\documentclass[11pt]{article}
\pdfoutput=1

\usepackage{jheppub}
\usepackage{amssymb,amsfonts,amsmath,amsthm}
\usepackage{mathrsfs}
\usepackage{bbold}
\usepackage{tikz}
\usetikzlibrary{shapes.geometric, arrows}
\usetikzlibrary{shapes.geometric,calc}
\usetikzlibrary{shapes,arrows,chains}
\usetikzlibrary{decorations.markings}
\usetikzlibrary{decorations.pathmorphing}
\usetikzlibrary{shapes.multipart}
\tikzset{snake it/.style={decorate, decoration=snake}}

\makeatletter
\newcommand{\Vast}{\bBigg@{4.75}}
\makeatother

\definecolor{darkred}{rgb}{0.9, 0.25, 0}
\definecolor{darkblue}{rgb}{0.188, 0.478, 0.9}

\newcommand{\be}{\begin{equation}}
\newcommand{\ee}{\end{equation}}
\newcommand{\bea}{\begin{eqnarray}}
\newcommand{\eea}{\end{eqnarray}}

\newcommand{\CA}{\mathcal{A}}

\newcommand{\CC}{\mathcal{C}}

\newcommand{\CI}{\mathcal{I}}

\newcommand{\CN}{\mathcal{N}}
\newcommand{\CM}{\mathcal{M}}

\newcommand{\CV}{\mathcal{V}}

\newcommand{\lr}{\left (}
\newcommand{\rr}{\right )}
\newcommand{\ls}{\left [}
\newcommand{\rs}{\right ]}

\newcommand\qt\tau

\newcommand{\p}{\partial}

\renewcommand{\tilde}[1]{\widetilde{#1}}

\makeatletter
\renewcommand{\@seccntformat}[1]{\csname the#1\endcsname.\,\,}
\makeatother
\allowdisplaybreaks

\let \savenumberline \numberline
\def \numberline#1{\savenumberline{#1.}}

\makeatletter
\def\@fpheader{\relax}
\makeatother

\def\bea{\begin{eqnarray}}
\def\eea{\end{eqnarray}}

\usetikzlibrary{shapes,arrows,chains}
\usetikzlibrary{decorations.markings}
\usetikzlibrary{decorations.pathmorphing}
\usetikzlibrary{shapes.multipart}
\tikzset{snake it/.style={decorate, decoration=snake}}

\title{\ \vspace{1.6cm} \\
KLT Factorization of Nonrelativistic String Amplitudes}
\author[a]{Ziqi Yan}
\author[b]{and Matthew Yu}
\emailAdd{ziqi.yan@su.se}
\emailAdd{myu@perimeterinstitute.ca}
\affiliation[a]{
Nordita, KTH Royal Institute of Technology and Stockholm University\\
Hannes Alfv\'{e}ns v\"{a}g 12, SE-106 91 Stockholm, Sweden}
\affiliation[b]{
Perimeter Institute for Theoretical Physics\\
31 Caroline St N, Waterloo, ON N2L 2Y5, Canada}
\abstract{We continue our study of the Kawai-Lewelle-Tye (KLT) factorization of winding string amplitudes in \cite{Gomis:2021ire}.  
In a toroidal compactification, amplitudes for winding closed string states factorize into products of amplitudes for open strings ending on an array of D-branes localized in the compactified directions; the specific D-brane configuration is determined by the closed string data.
In this paper, we study a zero Regge slope limit of the KLT relations between winding string amplitudes. 
Such a limit of string theory requires a critically tuned Kalb-Ramond field in the compact directions, and leads to a self-contained corner called nonrelativistic string theory. 
This theory is unitary, ultraviolet complete, and its string spectrum and spacetime S-matrix satisfy nonrelativistic symmetry. 
Moreover, the asymptotic states in nonrelativistic string theory necessarily carry nonzero windings. 
First, starting with relativistic string theory, we construct a KLT factorization of amplitudes for winding closed strings in the presence of a critical Kalb-Ramond field. 
Then, in the zero Regge limit, we uncover a KLT relation for amplitudes in nonrelativistic string theory. 
Finally, we show how such a relation can be reproduced from first principles in a purely nonrelativistic string theory setting. 
We will also discuss connections between these results and amplitudes of string theory in the discrete light cone quantization (DLCQ), a method that is relevant for Matrix theory.
}
\begin{document}

\maketitle
\vfill\eject

\section{Introduction}

In string theory, the tree-level closed string amplitudes factorize into a sum of quadratic products of open string amplitudes. This property is the well-known Kawai-Levellen-Tye (KLT) relation \cite{Kawai:1985xq}, which is a consequence of the factorization of complex integrals in closed string amplitudes into contour integrals in open string amplitudes. In the field theory limit, the KLT relations induce a factorization of graviton amplitudes into Yang-Mills amplitudes. Moreover, the KLT relations are also connected to the later discovered Bern-Carrasco-Johansson (BCJ) relations among amplitudes in gauge theory \cite{Bern:2008qj}. Such relations in both string theory and field theory not only bring simplifications in amplitude calculations, but also reveal intriguing intimacies between gravity and Yang-Mills theories, see also \cite{BjerrumBohr:2010hn,Cachazo:2014xea,He:2016mzd,Bern:1998sv}.

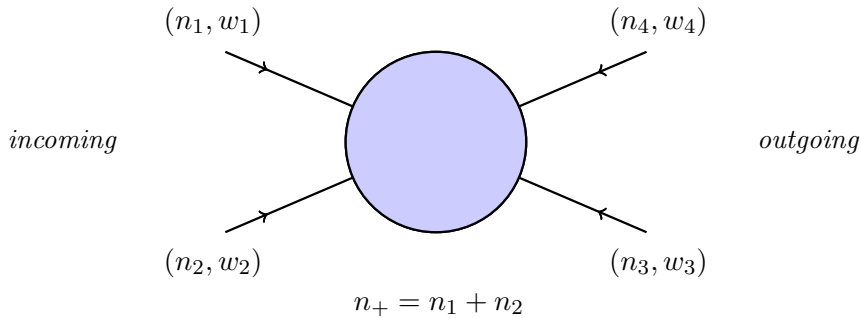
\begin{figure}[b!]
    \centering
    \begin{tikzpicture}[scale=0.8]
        \draw[line width = .3mm] (0,0) circle (1.5);
        \draw[decoration={markings, mark=at position .1 with
 {\arrow{>}}}, postaction={decorate}, decoration={markings, mark=at position .9 with
 {\arrow{<}}}, postaction={decorate},line width=.3mm] (-3.5,1.5) -- (3.5,-1.5);
        \draw[decoration={markings, mark=at position .1 with
 {\arrow{>}}}, postaction={decorate}, decoration={markings, mark=at position .9 with
 {\arrow{<}}}, postaction={decorate},line width=.3mm] (-3.5,-1.5) -- (3.5,1.5);
        \draw[fill=blue!20,line width = .3mm] (0,0) circle (1.5);
        \draw (-3.7,2) node {$(n_1, w_1)$};
        \draw (-3.7,-2) node {$(n_2, w_2)$};
        \draw (3.7,2) node {$(n_4, w_4)$};
        \draw (3.7,-2) node {$(n_3, w_3)$};
        \draw (-6.2,0) node {\scalebox{0.95}{\emph{incoming}}};
        \draw (6.2,0) node {\scalebox{0.95}{\emph{outgoing}}};
\end{tikzpicture}
\vspace{-6mm}
$$
    n_+ = n_1 + n_2
$$
\vspace{-9mm}
\caption{Closed string amplitude  with momentum   and winding  data  $(n_i,w_i)$ obeying      conservation laws $n_1 + n_2 + n_3 + n_4 = 0$ and $w_1 + w_2 + w_3 + w_4 = 0$\,.  Here, $n_1\, , n_2> 0$ and $n_3\, , n_4 \leq 0$,     splitting the strings into  incoming and  outgoing states. The integer $n_+$ is the  total incoming momentum.} 
\label{fig:fpcsa}
\end{figure}

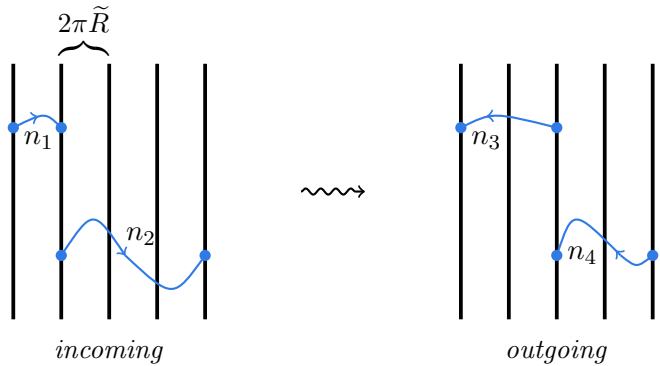
\begin{figure}[t!]
\centering
\begin{tikzpicture}[thick, scale=0.85]
    \draw[black,line width= .5mm] (-.5-3,-2) -- (-.5-3,2) ;
    \draw[black,line width= .5mm] (0.25-3,-2) -- (0.25-3,2) ;
    \draw[black,line width= .5mm] (1-3,-2) -- (1-3,2) ;
    \draw[black,line width= .5mm] (1.75-3,-2) -- (1.75-3,2) ;
    \draw[black,line width= .5mm] (2.5-3,-2) -- (2.5-3,2) ;
    \tikzstyle{vertex}=[circle,fill=darkblue!100, minimum size=1.5pt,inner sep=1.5pt]
    \node[vertex]at ( -3.5 , 1) {};
    \node[vertex]at ( -2.75 , 1) {};
    \node[vertex]at ( -2.75 , -1) {};
    \node[vertex]at ( -.5 , -1) {};
    \draw[darkblue, decoration={markings, mark=at position 0.5 with
        {\arrow{>}}}, postaction={decorate}] (-3.5,1).. controls (-3,1.25) .. (-2.75,1);
    \draw[darkblue, decoration={markings, mark=at position 1 with
        {\arrow{>}}}, postaction={decorate}] (-2.75,-1).. controls (-2.25,-.25) .. (-1.75,-1);
    \draw[darkblue] (-1.75,-1).. controls (-1,-1.7) .. (-.5,-1);
    \draw[->,decorate,decoration={snake,amplitude=.4mm,segment length=2mm,post length=1mm}] (1,0) -- (2, 0);
    \draw[black,line width= .5mm] (-.5-3+7,-2) -- (-.5-3+7,2) ;
    \draw[black,line width= .5mm] (0.25-3+7,-2) -- (0.25-3+7,2) ;
    \draw[black,line width= .5mm] (1-3+7,-2) -- (1-3+7,2) ;
    \draw[black,line width= .5mm] (1.75-3+7,-2) -- (1.75-3+7,2) ;
    \draw[black,line width= .5mm] (2.5-3+7,-2) -- (2.5-3+7,2) ;
    \tikzstyle{vertex}=[darkblue, circle,fill=darkblue!100, minimum size=1.5pt,inner sep=1.5pt]
    \node[vertex]at (-3.5+7, 1) {};
    \node[vertex]at (-2+7, 1) {};
    \node[vertex]at (-2+7, -1) {};
    \node[vertex]at (-.5+7, -1) {};
    \draw[darkblue, decoration={markings, mark=at position 0.35 with
        {\arrow{<}}}, postaction={decorate}] (-3.5+7,1).. controls (-3+7,1.25) .. (-2+7,1);
    \draw[darkblue,decoration={markings, mark=at position 1 with
        {\arrow{<}}}, postaction={decorate}] (-2+7,-1).. controls (-1.75+7,-.25) .. (-1+7,-1);
    \draw[darkblue] (-1+7,-1).. controls (-.75+7,-1.2) .. (-.5+7,-1);
    \node at (-3.1,0.8) {$n_1$};
    \node at (-3.1+7,0.8) {$n_3$};
    \node at (-1.5,-0.7) {$n_2$};
    \node at (-1.5+6.9,-1) {$n_4$};
    \node at (-2,-2.5) {\scalebox{0.95}{\emph{incoming}}};
    \node at (5,-2.5) {\scalebox{0.95}{\emph{outgoing}}};
    \node at (-2.4,2.15) {\scalebox{0.95}{$\overbrace{\phantom a}$}};
    \node at (-2.4,2.7) {\scalebox{0.95}{$2\pi \tilde{R}$}};
\end{tikzpicture}%
\caption{A four-point open string amplitude with $n_1 = 1$\,, $n_2 = 3$\,, $n_3 = -2$\,, and $n_4 = -2$\,. The vertical lines represent an array of equidistantly separated D-branes localized along the compact circle of radius $R$\,. The seemingly noncompact direction along which the D-branes are localized is understood to have an $x \sim x + 2 \pi R$ identification.
The distance between consecutive D-branes in the array is $2\pi\tilde{R}$\,, where $\tilde{R} = \alpha' / R$ is the radius of the T-dual circle. The blue wavy lines represent fundamental strings with their ends residing on different D-branes. During the scattering process, the two incoming strings join into one at the second D-brane from the left; then, the single intermediate string splits at the third D-brane into two outgoing strings. The open strings may also wind around the full compact circle for multiple times. There are $n_+ + 1 = 5$ D-branes for this choice of closed string quantum numbers.}
\label{fig:osamp}
\end{figure}

Recently, in \cite{Gomis:2021ire}, the KLT relations are generalized to amplitudes that involve winding string states,
which are responsible for some of the salient features of string theory such as T-duality. For simplicity and concreteness, we will focus on the compactification over a spacelike circle, and consider a factorization of the amplitude that describes an $\CN$-point scattering between closed string states that carry both winding and momentum along the compactified direction. This closed string amplitude is expressed by a quadratic product of amplitudes for open strings that end on an array of D-branes, whose configuration is determined by the closed string data as follows: These D-branes are localized in the compact circle, and are equally separated by a distance that is T-dual to the circumference of the compact circle. The total number of D-branes is determined by the total incoming momentum of the closed string scattering in the compact circle. A closed string state that carries a winding number $w$ and a Kaluza-Klein (KK) momentum number $n$ is mapped to a winding open string state in the KLT relation. This corresponding open string state describes an open string that winds $w$ times around the spatial circle and, in addition, traverses $n-1$ D-brane. Here, $n$ is associated with the fractional part of the open string winding number, with the open string ending at two different D-branes. In other words, the closed string winding and momentum in the compact circle are mapped to the integer and fractional open string windings, respectively. Finally, the conservation of closed string momenta in the compact circle is realized by open strings rejoining and splitting on D-branes. The above configurations of closed and open string scatterings in KLT relations between winding string amplitudes are illustrated by a four-point scattering process in Fig.~\ref{fig:fpcsa} and Fig.~\ref{fig:osamp}.~\footnote{See \cite{Gomis:2021ire} for discussions on a T-dual interpretation of open strings ending on the D-brane array illustrated in Fig.~\ref{fig:osamp}. After performing the T-duality transformation, the open strings live on a stack of spacetime-filling D-branes in the presence of Wilson lines.} See \S\ref{sec:rklt} for a more thorough review.

The standard KLT relation for string amplitudes without winding has a well-defined field theory limit at $\alpha' \rightarrow 0$\,, with $\alpha'$ the Regge slope. This limit leads to relations between gravitational and Yang-Mills amplitudes. In contrast, for scatterings that involve winding string states, such a field theory limit does not exist anymore, as all the winding states are gapped out.~\footnote{Also see the recent work \cite{Li:2021yfk} on a field theory limit for KK string amplitudes on a spacetime compactification, where the winding is set to zero.} However, there still exists a non-singular zero $\alpha'$ limit of winding string amplitudes, provided that a Kalb-Ramond $B$-field along the compact circle is present. In this limit of string theory, the $B$-field is fined tuned to its critical value such that it cancels the string tension \cite{Klebanov:2000pp, Gomis:2000bd, Danielsson:2000gi}.~\footnote{Historically, this limit is known as the noncommutative open string (NCOS) limit \cite{Seiberg:2000ms,Gopakumar:2000na}.} As a result, part of the states in the string spectrum are decoupled, and all the remaining physical states carry nonzero windings and satisfy a Galilean dispersion relation. This resulting theory is a string theory that has a unitary and ultraviolet (UV) finite spacetime S-matrix that enjoys nonrelativistic symmetry. We refer to this theory as \emph{nonrelativistic string theory} \cite{Gomis:2000bd}. Naturally, KLT relations for nonrelativistic strings can be uncovered by taking a limit of the KLT relations for winding string amplitudes in relativistic string theory. Establishing such KLT relations is among the first steps of unraveling new structures of amplitudes in nonrelativistic string theory. 

It is also known that nonrelativistic string theory is related to relativistic string theory in the discrete light cone quantization (DLCQ) via a T-duality transformation along the compact circle \cite{Gomis:2000bd, Danielsson:2000gi, Bergshoeff:2018yvt, Kluson:2018vfd, Bergshoeff:2019pij, Gomis:2020izd}.~\footnote{Also see \cite{Harmark:2017rpg, Kluson:2018egd, Harmark:2019upf, Harmark:2018cdl} for related works on null reductions.} The DLCQ of quantum field theories (QFTs) and string theory are important for various nonperturbative methods in quantum chromodynamics \cite{Brodsky1985} and Matrix theory \cite{Banks:1996vh}, respectively. The zero $\alpha'$ limit of winding string amplitudes is therefore also relevant for Matrix string theory \cite{Motl:1997th, Banks:1996my,Dijkgraaf:1997vv}.

\begin{figure}[t!]
\centering
\begin{tikzpicture}[thick, scale=0.85]
    \draw[black,line width= .5mm] (-.5-3,-2) -- (-.5-3,2) ;
    \node at (-.5-3,2.4) {\scalebox{0.9}{$V_{1}$}};
    \draw[black,line width= .5mm] (0.25-3,-2) -- (0.25-3,2) ;
    \node at (0.25-3,2.4) {\scalebox{0.9}{$V_{2}$}};
    \draw[black,line width= .5mm] (1-3,-2) -- (1-3,2) ;
    \node at (1-3,2.4) {$V_{3}$};
    \draw[black,line width= .5mm] (1.75-3,-2) -- (1.75-3,2) ;
    \node at (1.75-3,2.4) {\scalebox{0.9}{$V_{4}$}};
    \draw[black,line width= .5mm] (2.5-3,-2) -- (2.5-3,2) ;
     \node at (2.5-3,2.4) {\scalebox{0.9}{$V_{5}$}};
    \tikzstyle{vertex}=[circle,fill=darkblue!100, minimum size=1.5pt,inner sep=1.5pt]
    \node[vertex]at ( -3.5 , 1) {};
    \node[vertex]at ( -2.75 , 1) {};
    \node[vertex]at ( -2.75 , -1) {};
    \node[vertex]at ( -.5 , -1) {};
    \draw[darkblue, decoration={markings, mark=at position 0.5 with
        {\arrow{>}}}, postaction={decorate}] (-3.5,1).. controls (-3,1.25) .. (-2.75,1);
    \draw[darkblue, decoration={markings, mark=at position 1 with
        {\arrow{>}}}, postaction={decorate}] (-2.75,-1).. controls (-2.25,-.25) .. (-1.75,-1);
    \draw[darkblue] (-1.75,-1).. controls (-1,-1.7) .. (-.5,-1);
    \draw[->,decorate,decoration={snake,amplitude=.4mm,segment length=2mm,post length=1mm}] (1,0) -- (2, 0);
    \tikzstyle{vertex}=[circle,fill=black!70, minimum size=1pt,inner sep=.75pt]
    \node[vertex]at ( -1-3.5 , 0) {};
    \node[vertex]at ( -1-3.5+.2 , 0) {};
    \node[vertex]at ( -1-3.5+.4 , 0) {};
    \tikzstyle{vertex}=[circle,fill=black!70, minimum size=1pt,inner sep=.75pt]
    \node[vertex]at ( -1-3.5+7 , 0) {};
    \node[vertex]at ( -1-3.5+.2 +7, 0) {};
    \node[vertex]at ( -1-3.5+.4+7 , 0) {};
    \tikzstyle{vertex}=[circle,fill=black!70, minimum size=1pt,inner sep=.75pt]
    \node[vertex]at ( 0 , 0) {};
    \node[vertex]at ( 0+.2 , 0) {};
    \node[vertex]at ( 0+.4 , 0) {};
    \tikzstyle{vertex}=[circle,fill=black!70, minimum size=1pt,inner sep=.75pt]
    \node[vertex]at ( 0+7 , 0) {};
    \node[vertex]at ( 0+.2+7 , 0) {};
    \node[vertex]at ( 0+.4+7 , 0) {};
    \draw[black,line width= .5mm] (-.5-3+7,-2) -- (-.5-3+7,2) ;
     \node at (-.5-3+7,2.4) {\scalebox{0.9}{$V_{1}$}};
    \draw[black,line width= .5mm] (0.25-3+7,-2) -- (0.25-3+7,2) ;
     \node at (0.25-3+7,2.4) {\scalebox{0.9}{$V_{2}$}};
    \draw[black,line width= .5mm] (1-3+7,-2) -- (1-3+7,2) ;
    \node at (1-3+7,2.4) {\scalebox{0.9}{$V_{3}$}};
    \draw[black,line width= .5mm] (1.75-3+7,-2) -- (1.75-3+7,2) ;
    \node at (1.75-3+7,2.4) {\scalebox{0.9}{$V_{4}$}};
    \draw[black,line width= .5mm] (2.5-3+7,-2) -- (2.5-3+7,2);
    \node at (2.5-3+7,2.4) {\scalebox{0.9}{$V_{5}$}};
    \tikzstyle{vertex}=[darkblue, circle,fill=darkblue!100, minimum size=1.5pt,inner sep=1.5pt]
    \node[vertex]at (-3.5+7, 1) {};
    \node[vertex]at (-2+7, 1) {};
    \node[vertex]at (-2+7, -1) {};
    \node[vertex]at (-.5+7, -1) {};
    \draw[darkblue, decoration={markings, mark=at position 0.35 with
        {\arrow{<}}}, postaction={decorate}] (-3.5+7,1).. controls (-3+7,1.25) .. (-2+7,1);
    \draw[darkblue,decoration={markings, mark=at position 1 with
        {\arrow{<}}}, postaction={decorate}] (-2+7,-1).. controls (-1.75+7,-.25) .. (-1+7,-1);
    \draw[darkblue] (-1+7,-1).. controls (-.75+7,-1.2) .. (-.5+7,-1);
    \node at (-3.1,0.8) {$n_1$};
    \node at (-3.1+7,0.8) {$n_3$};
    \node at (-1.5,-0.7) {$n_2$};
    \node at (-1.5+6.9,-1) {$n_4$};
    \node at (-2,-2.5) {\scalebox{0.95}{\emph{incoming}}};
    \node at (5,-2.5) {\scalebox{0.95}{\emph{outgoing}}};
\end{tikzpicture}%
\caption{The $s$-th D-brane in the array carries an electric gauge potential $V_s = s B / R$\,, and the D-branes are separated by $2\pi \tilde{R}$.}
\label{fig:osampwithpotential}
\end{figure}
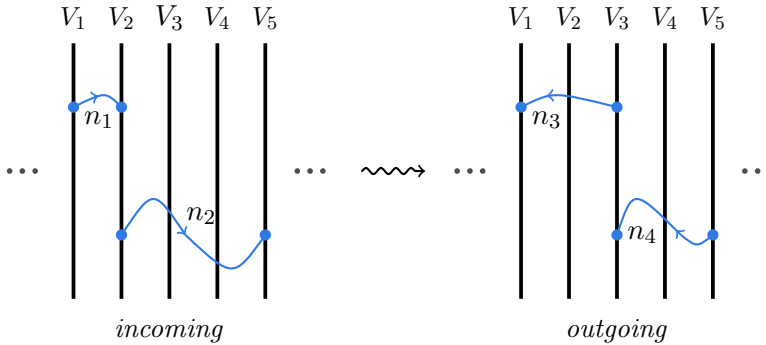

To facilitate such a zero Regge slope limit of KLT relations for relativistic string amplitudes, we first study in relativistic string theory a generalization of KLT factorization of amplitudes for winding closed string states, coupled to a constant $B$-field in the compact circle.
This $B$-field has the effect of shifting the energy of both closed and open string states, with the shift being proportional to the winding. 
Naively, the factorization gains the interpretation as a product of amplitudes for open strings that end on the same array of D-branes as illustrated in Fig.~\ref{fig:osamp}. However, since the closed string KK number is mapped to the fractional part of the open string winding in the compact circle, the shift of the open string energy due to the $B$-field also contains an extra contribution from this fractional winding, which is absent on the closed string side of the KLT relation. This mismatch between energies of closed and open string states requires an additional structure on the D-brane array: each D-brane now carries a different electric gauge potential, whose value is determined by the closed string momenta in the compactified direction. See Fig.~\ref{fig:osampwithpotential} for an illustration of this modified D-brane configuration, which resembles a series of capacitors with a constant electric field between the plates. This electric field is only experienced by open strings that end on different D-branes. See relevant discussions in \S\ref{sec:wkltbf} for this generalized KLT relations in a background $B$-field.     

After developing this new KLT relation in $B$-field for relativistic string amplitudes, we will finally set the $B$-field to its critical value and apply the zero Regge slope limit. As we have indicated earlier, this procedure leads to a KLT relation in nonrelativistic string theory. The amplitudes of asymptotic closed string states, which necessarily carry nonzero windings in nonrelativistic string theory, factorize into a product of amplitudes for open strings that end on a D-brane configuration that is determined by the zero $\alpha'$ limit of the one illustrated in Fig.~\ref{fig:osampwithpotential}. Recall that the consecutive D-branes in the array are separated by a distance $2 \pi \tilde{R}$ that is T-dual to the circumference of the compact circle, and this distance becomes zero after sending $\alpha'$ to zero. Consequently, the D-branes become coincident. The values of gauge potentials $V_s$ are unaffected by the limit, but they now become the diagonal entries of a single gauge field that is in the diagonal subgroup $U(1)^{n_+ + 1}$ of $U(n_+ \! + 1)$\,. See \S\ref{sec:zalwkltr} for further details.   

The same KLT relation in nonrelativistic string theory can also be reproduced by using a first principles approach, independent of the zero Regge slope limit. In \cite{Gomis:2000bd}, a self-consistent worldsheet QFT that describes nonrelativistic string theory was formulated. In flat spacetime, this worldsheet theory enjoys a (string)-Galilean invariant global symmetry. Due to recent analysis of this worldsheet theory and improved understanding of non-Riemannian geometries, these years have witnessed a growing interest in nonrelativistic string theory in curved spacetime, see \emph{e.g.} \cite{Andringa:2012uz, Bergshoeff:2018yvt}.~\footnote{Also see \cite{Bergshoeff:2019pij} for more modern review.} It has been shown that this string theory, being UV complete on its own, provides a quantization of the so-called (torsional) string Newton-Cartan geometry that is non-Riemannian, akin to how relativistic string theory provides a quantization of Einstein gravity.~\footnote{See \cite{Gomis:2019zyu, Yan:2019xsf, Gallegos:2019icg} for studies of Weyl anomalies and \cite{Harmark:2018cdl, Harmark:2019upf, Gallegos:2020egk, Bergshoeff:2021bmc, Yan:2021lbe, Bidussi:2021ujm} for torsional extensions. Supersymmetric generalizations of the background geometries have been considered in \cite{Gomis:2005pg, Park:2016sbw, Blair:2019qwi, Bergshoeff:2021tfn}. For perspectives from Double Field Theory, see, \emph{e.g}, \cite{Ko:2015rha, Park:2016sbw, Morand:2017fnv, Blair:2019qwi, Gallegos:2020egk, Blair:2020gng}.} Similar analysis has also been applied to nonrelativistic open strings, from which a Dirac-Born-Infeld action that describes the low-energy dynamics of D-branes in nonrelativistic string theory is derived \cite{Gomis:2020fui}.~\footnote{Also see the companion paper \cite{ddanst} for explorations of dual D-brane actions in nonrelativistic string theory.}

In parallel to the above successes in nonrelativistic string theory, the amplitude analysis remains under-explored other than the early works \cite{Gomis:2000bd, Danielsson:2000gi} on nonrelativistic closed string amplitudes. In \S\ref{sec:kltfnsa}, we will use the worldsheet theory that is originally proposed in \cite{Gomis:2000bd} to derive the KLT relation in nonrelativistic string theory, and corroborates the factorization that we derive from a limit of the relativistic KLT relation in \S\ref{sec:zalwkltr}. Finally, in \S\ref{sec:stdlcq}, we will study the T-dual interpretation of this KLT relation and its connection to the DLCQ of string theory. As a further demonstration of the techniques developed through the paper, in Appendix \ref{app:ola}, we apply these techniques to compute one-loop amplitudes in nonrelativistic open string theory, generalizing properties of nonrelativistic closed string amplitudes in \cite{Gomis:2000bd, Danielsson:2000gi} to open strings. We conclusion the paper in \S\ref{sec:concl}.

\section{KLT Factorizations in Relativistic String Theory} \label{sec:kltfwsdb}

We start with a review of the KLT factorization of closed string amplitudes into open string amplitudes in toroidal spacetime compactification, in relativistic string theory. We will follow closely \cite{Gomis:2021ire} for this review. For simplicity, we focus on the case where a single spacial direction is compactified. We consider scattering between asymptotic closed string states that carry both winding and momentum in the compactified direction. In the KLT relations, the closed string winding and Kaluza-Klein numbers are mapped respectively to the integer and fractional windings of open strings, which end on an array of D-branes localized equidistantly along the compactified direction. The open string amplitudes involved in the winding KLT relation are therefore associated with open strings ending on such a D-brane configuration determined by the closed string quantum numbers. Then, we move on to the construction of a winding KLT relation in the presence of a Kalb-Ramond field, which is important for us to perform at the end a non-singular zero Regge slope limit. Such a low energy limit of scattering amplitudes between winding states does \emph{not} lead to any field theory regime; instead, we show  that the resulting amplitudes are physical observables in the so-called nonrelativistic string theory, a unitary and UV-complete string theory that has a string spectrum with a Galilean invariant dispersion relation.    

\subsection{KLT Relations for Winding String Amplitudes} \label{sec:rklt}

Consider relativistic string theory described by a sigma model that maps a two-dimensional worldsheet $\Sigma$ to a $d$-dimensional spacetime manifold $\CM$\,. We will focus on bosonic strings with $d=26$\,. The worldsheet fields $X^\mu$\,, $\mu = 0,\, 1, \, \cdots, \, d-1$ play the role of spacetime coordinates. We compactify $X^1$ over a circle of radius $R$\,, such that $X^1 \sim X^1 + 2\pi R$\,.
We therefore split the target space into a two-dimensional longitudinal and ($d-2$)-dimensional transverse sector, carrying the index $A = 0, 1$ and $A' = 2, \cdots, d-1$\,, respectively.  
We parametrize the worldsheet $\Sigma$\,, which is isomorphic to the complex plane, by the complex coordinates $z$ and $\bar{z}$\,. Dividing the spacetime coordinates $X^\mu$ into the left- and right-moving parts, we write
\be
    X^\mu (z,\, \bar{z}) = X^\mu_\text{L} (z) + X^\mu_\text{R} (\bar{z})\,.
\ee
For a closed string state that carries a KK momentum number $n$ and winding number $w$ in $X^1$, with $n, w \in \mathbb{Z}$\,, we have the general form of the closed string vertex operator 
\be \label{eq:vo}
    e^{i \, \pi \, w \, \hat{n}} \, :\! \p_z^{k_1} \! X^{\mu_1}_{\text{L}}  \, \cdots \, \p_{z}^{k_p} \! X^{\mu_p}_{\text{L}} \,
    \p_{\overline{z}}^{\ell_1} \! X^{\nu_1}_{\text{R}} \, \cdots \, \p_{\overline{z}}^{\ell_p} \! X^{\nu_p}_{\text{R}} \,\, e^{i K_\text{L} \cdot X_\text{L} + i K_\text{R} \cdot X_\text{R}} \!:\,,
\ee
where 
\be \label{eq:klr}
    K_{\text{L}} = \lr \varepsilon\,, \, \frac{n}{R} - \frac{wR}{{\alpha}'}\,, \, k_{A'} \rr,
        \qquad%
    K_{\text{R}} = \lr \varepsilon\,, \, \frac{n}{R} + \frac{wR}{{\alpha}{}'}\,, \, k_{A'} \rr.
\ee
Here, ${\alpha}{}'$ is the Regge slope. General BRST vertex operators can then be constructed using the building blocks in \eqref{eq:vo} \cite{Gomis:2021ire}. We define $\varepsilon$ as the energy and $k_{A'}$ the transverse momentum. We also introduce the operator 
\be
    \hat{n} = \frac{R}{2\pi \alpha'} \oint_\CC \bigl( dz \, \p_z X_\text{L} - d\bar{z} \, \p_{\bar{z}} X_\text{R} \bigr),
\ee
whose eigenvalues are the KK number $n$\,. The phase factor $e^{i \, \pi \, w \, \hat{n}}$ in \eqref{eq:vo} is the \emph{cocycle} factor that is introduced to ensure that the vertex operators commute with each other \cite{Polchinski:1998rq}.  

The simplest nontrivial case we will consider is the tree-level four-tachyon scattering on a sphere. For tachyonic states, we have the dispersion relation $K_\text{L}^2 = K_\text{R}^2 = 4 / \alpha'$ and the level-matching  $n \, w = 0$\,. Locations of the vertex operators on the complex plane are denoted by $z_i$\,, $i = 1, \cdots, 4$\,. We use the M\"{o}bius transformation to fix the locations of three vertices, with $z_1 = 0$\,, $z_3 = 1$\,, and $z_4 = \infty$\,. The closed string amplitude is of the Virasoro-Shapiro-type, 
\begin{align} \label{eq:csa}
\begin{split}
    {\CM}^{(4)}_\text{c} & = {C}(1,2,3,4) \\[2pt]
    & \quad \times \int_\mathbb{C} d^2 z \, z^{\frac{1}{2} \, {\alpha}{}' K_{\text{L}1} \cdot K_{\text{L}2}} \, \bar{z}^{\frac{1}{2} \, {\alpha}{}' K_{\text{R}1} \cdot K_{\text{R}2}} \, (1-z)^{\frac{1}{2} \, {\alpha}{}' K_{\text{L}2} \cdot K_{\text{L}3}} \, (1-\bar{z})^{\frac{1}{2} \, {\alpha}{}' K_{\text{R}2} \cdot K_{\text{R}3}}\,,
\end{split}
\end{align}
where $z = z_2$ and
\be
    {C}(1,2,3,4) = \exp \! \lr i \pi {\sum}_{\substack{i,j=1\\i<j}}^4 \, n_i \, w_j \rr
\ee
comes from the cocycle factors and ensure that the vertex operators commute with each other. We have omitted the Dirac deltas that impose various conservation laws for both momenta and windings in the amplitude, in addition to other prefactors that are unimportant when the KLT relation is concerned. The closed string amplitude \eqref{eq:csa} factorizes as \cite{Gomis:2021ire}
\be \label{eq:kltws}
    {\CM}^{(4)}_\text{c} = - \, {C}(1,2,3,4) \, {\CM}_\text{L} (2,1,3,4) \, \sin \lr \tfrac{1}{2} \, \pi \, {\alpha}{}' \, K_{\text{L}1} \cdot K_{\text{L}2} \rr {\CM}_\text{R} (1,2,3,4)\,,
\ee
where
\begin{subequations} \label{eq:mlr}
\begin{align}
    \CM_\text{L} (2,1,3,4) & = \int_{-\infty}^0 dy \, (-y)^{\frac{1}{2} {\alpha}{}' K_{\text{L}1} \cdot K_{\text{L}2}} \, (1-y)^{\frac{1}{2} {\alpha}{}' K_{\text{L}2} \cdot K_{\text{L}3}}\,, \label{eq:ml} \\[2pt]
    \CM_\text{R} (1,2,3,4) & = \int_0^1 dy \, y^{\frac{1}{2} {\alpha}{}' K_{\text{R}1} \cdot K_{\text{R}2}} \, (1-y)^{\frac{1}{2} {\alpha}{}' K_{\text{R}2} \cdot K_{\text{R}3}}\,. \label{eq:mr}
\end{align}
\end{subequations}
Note that the KLT-like relation in \eqref{eq:kltws} is invariant under swapping the subscripts ``L" and ``R." The quantities $\CM_\text{L,\,R}$ can be interpreted as open string amplitudes. An intriguing subtlety arises in this interpretation: The closed string state in the compactified spacetime can carry both momentum $n/R$ and winding $w$ along the compact circle in $X^1$;~\footnote{Each individual tachyonic closed string state considered here cannot carry both winding and momentum in the compact circle, due to the level-matching condition $n \, w = 0$\,. However, in the same scattering process, it is possible to have some asymptotic closed string states with nonzero windings and others with nonzero momenta along the compact circle. Therefore, the four-tachyon amplitude still provides an interesting example that exhibits most salient features of KLT relations for winding string amplitudes. If, instead, the more general vertex operators of the form \eqref{eq:vo} are considered, then each individual closed string state can in general carry both winding and momentum in the compact circle. See \cite{Gomis:2021ire} for further discussions on insertions of general vertex operators.} in contrast, an open string state can only carry momentum or winding along $X^1$, but not both of them, depending on which boundary condition (Neumann or Dirichlet) is imposed. This observation implies that, in $\CM_\text{L,\,R}$ from \eqref{eq:mlr} with $K_\text{L,\,R}$ defined in \eqref{eq:klr}, we \emph{cannot} simultaneously interpret $n/R$ as momentum and $w$ as winding for any open string state. 

In \cite{Gomis:2021ire},  it is shown that a particular D-brane configuration is required for interpreting $\CM_\text{L,\,R}$ in \eqref{eq:mlr} as open string amplitudes. While the winding number of the $i$-th closed string corresponds to the winding number of the $i$-th open string obeying Dirichlet boundary conditions, the closed string KK momentum number is encoded in the D-brane configuration where open string amplitudes are defined. We elaborate on this D-brane configuration below.
Consider an array of D-branes that are localized in the compactified $X^1$-direction but extend in all the noncompact directions. Require that the $s$-th brane be located in $X^1$ at 
\be
    x_s = x_0 + s \, L \quad \text{mod} \quad 2\pi R\,,
        \qquad
    s \in \mathbb{Z}\,,
\ee
where $L = 2 \pi {\alpha}{}' / R$ is T-dual to the circumference of the compact circle. Consider the $i$-th open string, and let one of its ends be anchored on the $s$-th D-brane. After wrapping around the $X^1$ circle for $w_i$ times, we fix the other end of the open string on the $(s+n_i)$-th D-brane. Then, the total winding number of the open string is
\be \label{eq:twn}
    W_i = w_i + \frac{n_i L}{2\pi R}\,,
\ee
where $n_i \, L / (2\pi R)$ is generically fractional. Therefore, while $w_i$ corresponds to the integer winding number of the $i$-th open string, the closed string KK number $n_i$ for the $i$-th closed string is encoded by the fractional winding. The conservation law of the closed string momentum in $X^1$ is realized by open strings splitting and joining on the D-branes. The number of D-branes involved in the scattering process is given by $n_+ + 1$\,, where $n_+$ is the sum of all positive $n_i$\,, associated with the incoming closed strings. 
see Fig.~\ref{fig:fpcsa} and \ref{fig:osamp} for an illustration.
For the given cyclic ordering $(1,2,3,4)$ of the vertex operators on the boundary of a worldsheet disk, and after the rescalings, 
\be \label{eq:rs}
    {\alpha}{}' \rightarrow \frac{{\alpha}{}'}{4}\,,
        \qquad%
    R \rightarrow \frac{R}{4}\,,
        \qquad%
    L \rightarrow \frac{L}{4}\,,
\ee
the resulting four-point open string amplitude exactly reproduces $\CM_\text{R}$ in \eqref{eq:mr}. Analogously, the same interpretation in terms of open string amplitudes applies to $\CM_\text{L}$ in \eqref{eq:ml}, but with a different cyclic ordering $(2,1,3,4)$ and an opposite sign in front of $w_i$\,. In this sense, \eqref{eq:kltws} presents a factorization of closed string amplitudes into a quadratic product of open string amplitudes, with the open strings ending on a D-brane configuration that encodes the closed string winding and KK numbers. 

Finally, consider an $\CN$-point closed string amplitude $ {\CM}_\text{c}^\CN$ that is defined up to an overall coefficient and involves general vertex operators that consist of the ingredients in \eqref{eq:vo}. We use the M\"{o}bius transformation to fix the locations of three vertex operators on the complex plane, with $z_1 = 0\,, z_{\CN-1} = 1$ and $z_\CN = \infty$\,. The KLT factorization of $ {\CM}_\text{c}^\CN$ is
%
\cite{Gomis:2021ire}
\begin{align} \label{eq:rcsamp}
\begin{split}
    {\CM}_\text{c}^\CN & = (-1)^{\CN-3} \sum_{\rho,\,\sigma} {C} (1, \, \sigma, \, \CN-1, \, \CN) \\[-4pt]
    & \hspace{2cm} \times {\CM}_\text{L} \lr \rho, \,1, \, \CN-1, \, \CN \rr \,
    \, {\mathcal{S}}_\text{L}[\rho|\sigma]_{{K}_{\text{L}1}} \,
    {\CM}_\text{R} \lr 1, \, \sigma, \, \CN-1, \, \CN \rr,
\end{split}
\end{align}
where $\rho$ and $\sigma$ denote permutations of  the indices in $\{ 2, \cdots, \, \CN-2 \}$\,, and
\begin{align}\label{momentumkernel}
    \mathcal{S}_\text{L} [i_1\,, \dots, i_k | j_1\,, \cdots, j_k]^{}_{P}
    = \prod_{t=1}^k \sin \Bigl[ \tfrac{1}{2} \, \pi \, {\alpha}{}' \Bigl( P \cdot {K}_{\text{L}i_t} + \sum_{q>t}^k \theta(i_t, i_q) \, {K}_{\text{L}i_t} \cdot {K}_{\text{L}i_q} \Bigr) \Bigr]
\end{align}
is the momentum kernel. The kinematic data $K_{\text{L}i}$ and $K_{\text{R}i}$ are already defined in \eqref{eq:klr}, but now dressed up with the subscript $i$ that labels asymptotic string states in the scattering. Moreover, $(i_1 ,...,i_k)$ and $(j_1 ,...,j_k)$ are permutations of $\{1,...,k\}$\,. If the ordering of $i_t$ and $i_q$ is the opposite in the ordered sets $(i_1 ,... ,i_k)$ and $(j_1 ,... ,j_k)$, we set $\theta(i_t ,i_q) = 1$; otherwise, we set $\theta(i_t ,i_q) = 0$\,.
The cocycle factor takes the following form:
\be \label{eq:cfori}
    {C} (i_1, \dots, \, i_\CN) = \exp \lr i \pi {\sum}_{\substack{p,\,q=1\\p<q}}^\CN \, n_{i_p} w_{i_q} \rr,
\ee
which is required for the commutativity of vertex operators.
We also defined the $\CN$-point open string amplitudes as we have described earlier for the four-point case.
\be \label{eq:osamnr2}
    {\CM}_\text{L,\,R} (1, \dots, \CN) = \int_{0 < y_2 < \cdots < y_{\CN-1} < 1} dy_2 \, \cdots \, dy_{\CN-2}\, F_\text{L,\,R}  \, \prod_{\substack{i,j = 1\\i<j}}^{\CN-1} |y_i - y_j|^{\frac{{\alpha}{}'}{2} {K}_{\text{L,R}i} \cdot {K}_{\text{L,R}j}},
\ee
and similarly for open string amplitudes that involve different cyclic orderings of open string vertex operators inserted on the boundary of the worldsheet disk.
The factors $F_\text{L,\,R}$ come from contracting derivative terms in the open string vertex operators, and are single-valued functions that do not contribute any branch points.
For a four-point scattering, \eqref{eq:rcsamp} reduces to \eqref{eq:kltws}. Again, the KLT-like relation in \eqref{eq:rcsamp} is invariant under swapping ``L" and ``R." Other equivalent forms of the winding KLT relation \eqref{eq:rcsamp} can be found in \cite{Gomis:2021ire}.

\subsection{Winding KLT Relations in Kalb-Ramond Field} \label{sec:wkltbf}
 
For later applications to nonrelativistic string theory, we now consider modifications to the KLT factorization of winding string amplitudes in the presence of a constant Kalb-Ramond $B$-field in the longitudinal sector. To realize such a KLT relation, in addition to the D-brane configuration that we have discussed in \S\ref{sec:rklt}, we will show that the D-branes also need to carry electric gauge potentials that are determined by closed string quantum numbers.

We start with the sigma model that describes relativistic string theory in a constant $B$-field in the compact circle.
Consider a worldsheet $\Sigma$ with the coordinates $\sigma^\alpha = (\tau, \sigma)$\,, where $\tau = i \, \sigma^0$ is the imaginary time. We take $\tau \in \mathbb{R}$ and $\sigma \in [0, 2\pi]$\,. The worldsheet $\Sigma$ is mapped to a spacetime manifold $\CM$ with the longitudinal coordinates $X^A$, $A = 0,1$ and transverse coordinates $X^{A'}$, $A' = 2, \dots, d-1$\,. As before, $X^1$ is compactified over a circle of radius $R$\,, such that for a state that carries the winding number $w$\,, 
\be
    X^1 (\sigma + 2\pi) = X^1(\sigma) - 2 \pi R \, w\,,
        \qquad
    w \in \mathbb{Z}\,.
\ee
The sigma model is
\be \label{eq:nrsa0}
    S = \frac{1}{4\pi\alpha'} \int_\Sigma d^2 \sigma \left( \p_\alpha X^{\mu} \, \p^\alpha X_{\mu} + + i \, \epsilon^{\alpha\beta} \, \p_\alpha X^\mu \, \p_\beta X^\nu \, B_{\mu\nu} \right),
\ee
with
\be 
    {B}_{\mu\nu} = 
        \begin{pmatrix}
            B \, \epsilon_{AB} & \,\, 0 \\[2pt]
            0 & \,\, 0
        \end{pmatrix},
\ee
where $B$ denotes the magnitude of the $B$-field in the compact circle, and the Levi-Civita symbol $\epsilon^{\alpha\beta}$ is defined by $\epsilon^{\tau\sigma} = - \epsilon_{\sigma\tau} = 1$\,, while $\epsilon_{AB}$ is defined by $\epsilon_{01} = - \epsilon_{10} = 1$\,. The canonical momentum conjugate to $X^0$ is
\be
    K_0 = \frac{1}{2\pi \alpha'} \int_0^{2\pi} d\sigma \, \lr i \, \p_\tau X^0 + B \, \p_\sigma X^1 \rr = \varepsilon - \frac{w R}{\alpha'} \, B\,,
\ee
where
\be
    \varepsilon = \frac{i}{2\pi \alpha'} \int_0^{2\pi} d\sigma \, \p_\tau X^0
\ee
is the kinetic energy in the absence of any $B$-field. 
Therefore, the $B$-field in the compact direction shifts the kinetic energy, with the shift being proportional to both the $B$-field and winding. 
For a closed string state with winding number $w$ and KK number $n$\,, the associated dispersion relation picks up the same energy shift generated by the $B$-field, with
\be
    \lr \! \varepsilon - \frac{wR}{\alpha'} \, B \! \rr^2 - k_{A'} k_{A'} = \frac{n^2}{R^2} + \frac{2}{\alpha'} \lr N + \tilde{N} - 2 \rr,
\ee
where $N$ and $\tilde{N}$ are string excitation numbers for the left and right movers, respectively.
The factorization of winding closed string amplitudes is in form the same as \eqref{eq:rcsamp}, but with
\begin{subequations} \label{eq:cskbf}
\begin{align}
    K_{\text{L}i} = \lr \varepsilon - \frac{w_i R}{\alpha'} \, B\,, \, \frac{n_i}{R} - \frac{w_i R}{{\alpha}'}\,, \, k_{A'\!,\, i} \rr, \label{eq:klmB} \\[2pt]
    K_{\text{R}} = \lr \varepsilon - \frac{w_i R}{\alpha'} \, B\,, \, \frac{n_i}{R} + \frac{w_i R}{{\alpha}{}'}\,, \, k_{A'\!,\, i} \rr, \label{eq:krmB}
\end{align}
\end{subequations}
for the $i$-th string state.

In the above KLT factorization in the presence of a $B$-field, we have the same factor $\CM_{\text{L},\,\text{R}}$ defined by \eqref{eq:osamnr2}, but now with the new kinematic data in \eqref{eq:cskbf}. For such $\CM_{\text{L},\,\text{R}}$ to receive an interpretation as open string amplitudes, we continue to consider the same scattering in \S\ref{sec:rklt} that involves open strings ending on the D-brane array localized in $X^1$\,. For open strings, we take the worldsheet to be a strip with $\tau \in \mathbb{R}$ and $\sigma \in [0, \pi]$\,. The boundary $\p\Sigma$ of the worldsheet resides on $\sigma = 0, \pi$\,. The total winding number $W$ for an open string that wraps $w$ times around $X^1$ and ending on two different D-branes is given in \eqref{eq:twn}, with $W = w + n L / (2\pi R)$\,. After applying the rescaling \eqref{eq:rs}, the associated open string amplitude now takes the form of $\CM_{\text{R}}$ in \eqref{eq:osamnr2} but with 
\be \label{eq:krosb}
    K_{\text{R}} = \lr \varepsilon - \frac{W R}{\alpha'} \, B\,, \frac{WR}{\alpha'}\,, k_{A'} \rr.
\ee
Performing the rescaling \eqref{eq:rs}, and then using $L = 2 \pi \alpha' / R$\,, we find
\be \label{eq:krreb}
    K_\text{R} \rightarrow \lr \varepsilon - \frac{w R}{\alpha'} \, B -  \frac{n B}{R}\,, \frac{n}{R} + \frac{w R}{\alpha'}\,, k_{A'} \rr.
\ee
This is almost the same as \eqref{eq:krmB} but with an extra shift $n B / R$ of the energy. To compensate for this extra term, we are required to assign a constant electric potential $V_s$ to the $s$-th D-brane. The specific value of $V_s$ is determined as follows.
Consider an open string with the end at $\sigma = 0$ residing on the $s$-th and the other end at $\sigma = \pi$ residing on the $(s+n)$-th D-brane. It is convenient to write the boundary action in terms of the complex coordinate $z = e^{\tau + i \sigma}$ and its complex conjugate $\bar{z} = e^{\tau - i \sigma}$,
under which the worldsheet $\Sigma$ is mapped to the upper half-plane of $\mathbb{C}$\,: the boundary at $\sigma = 0$ is mapped to be the positive part of the real axis, and the boundary at $\sigma = \pi$ is mapped to be the negative part of the real axis. In this convention, the boundary action takes the form
\begin{align} \label{eq:bdrya}
    S_\text{bdry} & = i \int_{\p\Sigma} dy \, A_\mu \, \p_y X^\mu
    = i \int_0^\infty dy \, V_{s+n} \, \p_y X^0 + i \int_{-\infty}^0 dy \, V_{s} \, \p_y X^0\,.
\end{align}
Note that $y = e^\tau$ for $y \in (0, \infty)$\,, and $y = e^{\tau + i \pi}$ for $y \in (-\infty,0)$\,. In terms of $\tau$\,, we find
\be
    S_\text{bdry} = i \, (V_{s+n} - V_s) \int_{\mathbb{R}} d\tau \, \p_\tau X^0.
\ee
This boundary term contributes an extra shift in energy that is only experienced by open strings that end on different D-branes, with
\be \label{eq:sen}
    \varepsilon \rightarrow \varepsilon + V_{s+n} - V_n\,.
\ee
We require that
\be \label{eq:Vs}
    V_s = \frac{s L B}{2\pi\alpha'}\,,
        \qquad
    s \in \mathbb{Z}\,.
\ee
Performing the rescaling \eqref{eq:rs}, and then using $L = 2 \pi \alpha' / R$\,, we find
\be \label{eq:vsfinal}
    V_s = \frac{s B}{R}\,,
        \qquad%
    s \in \mathbb{Z}\,.
\ee
Taking the extra energy shift \eqref{eq:sen} into account, the kinematic data in \eqref{eq:krreb} becomes
\be \label{eq:korv}
    K_\text{R} = \lr \varepsilon - \frac{w R}{\alpha'} \, B -  \frac{n B}{R} + V_{s+n} - V_n\,, \frac{n}{R} + \frac{w R}{\alpha'}\,, k_{A'} \rr.
\ee
Plugging \eqref{eq:vsfinal} into \eqref{eq:korv}, we find that the same kinetic data in \eqref{eq:krmB} is recovered. Similar considerations also apply to $K_{\text{L}}$ in \eqref{eq:klmB}. 
Therefore, the amplitude for winding closed strings in an electric $B$-field lying along the compactified direction factorizes into a quadratic product of open string amplitudes, which involve open strings ending on an array of D-branes, each carrying a different electric potential.

Generally, the closed string metric and Kalb-Ramond fields are mapped to the effective metric that is seen by the open strings via the Seiberg-Witten map, together with a functional coupling that parametrizes spacetime noncommutativity \cite{Seiberg:1999vs}. In our case, the D-branes are transverse to the compactified direction, along which the $B$-field lies. Therefore, the $B$-field is orthogonal to the D-branes. In other words, the open strings satisfy Dirichlet boundary conditions in the direction of the $B$-field. Consequently, the effective field theories on the D-branes are not affected by the $B$-field and the Seiberg-Witten map trivializes. This is different from the situation in \cite{Seiberg:1999vs}, where Neumann boundary conditions are assumed along the $B$-field.

\subsection{A Zero \texorpdfstring{$\alpha'$}{alphap} Limit of Winding KLT Relations} \label{sec:zalwkltr}

Unlike the conventional KLT relations in a noncompactified spacetime, there is no field theory limit of the winding KLT relation \eqref{eq:rcsamp} by sending the Regge slope $\alpha'$ to zero. We focus on the following kinematic variable that shows up in \eqref{eq:rcsamp}:
\be \label{eq:kinv}
    \alpha' K_{\text{L}i} \cdot K_{\text{L}j} = \alpha' \ls - \varepsilon_i \, \varepsilon_j + \lr \frac{n_i}{R} - \frac{w_i R}{\alpha'} \rr \lr \frac{n_j}{R} - \frac{w_j R}{\alpha'} \rr + k_i^{A'} k_j^{A'} \rs.
\ee
In small $\alpha'$\,, \eqref{eq:kinv} becomes
\be \label{eq:daklkl}
    \alpha' \, K_{\text{L}i} \cdot K_{\text{L}j} \rightarrow \frac{1}{\alpha'} \Bigl[ w_i \, w_j \, R^2 + O (\alpha') \Bigr]\,,
\ee
which blows up in the $\alpha' \rightarrow 0$ limit and therefore leads to singular results. This is because any winding state drops out of the spectrum in the field theory limit. 
However, as we have preluded earlier, in the presence of an electric $B$-field that is tuned to its critical value with $B = -1$\,, there is a non-singular $\alpha' \rightarrow 0$ limit that leads to nonrelativistic string theory.

We start with introducing such a nonrelativistic limit at the level of the string sigma model in background fields,
\be \label{eq:hs}
    S = \frac{1}{4\pi\alpha'} \int_\Sigma d^2 \sigma \left( \p_\alpha X^{\mu} \, \p^\alpha X^{\nu} \, G_{\mu\nu} + i \, \epsilon^{\alpha\beta} \, \p_\alpha X^\mu \, \p_\beta X^\nu \, B_{\mu\nu} \right),
\ee
with
\be \label{eq:hgb2}
    {G}_{\mu\nu} = 
        \begin{pmatrix}
            \eta_{AB} & \,\, 0 \\[2pt]
            0 & \,\, \frac{\alpha'}{\alpha'_\text{eff}} \, \delta_{A'B'} 
        \end{pmatrix},
            \qquad%
    {B}_{\mu\nu} = 
        \begin{pmatrix}
            - \epsilon_{AB} & \,\, 0 \\[2pt]
            0 & \,\, 0
        \end{pmatrix}.
\ee
The coefficient $\alpha'_\text{eff}$ will become the effective Regge slope in nonrelativistic string theory. We continue to assume that $X^1$ is compactified along a circle of radius $R$\,.
Integrating in a pair of auxiliary fields $\lambda$ and $\bar{\lambda}$\,, the action \eqref{eq:hs} can be written equivalently as
\be \label{eq:lla}
    S = \frac{1}{4\pi\alpha'_\text{eff}} \int_\Sigma d^2 \sigma \lr \p_\alpha X^{A'} \, \p^\alpha X^{A'} + \lambda \, \bar{\p} \overline{X} + \bar{\lambda} \, \p X + \frac{\alpha'}{\alpha'_\text{eff}} \, \lambda \, \bar{\lambda} \rr.
\ee
We defined 
\be \label{eq:pbp}
    \p = i \, \p_\tau + \p_\sigma\,,
        \qquad%
    \bar{\p} = - i \, \p_\tau + \p_\sigma\,,
\ee
and the lightlike coordinates 
\be \label{eq:XbX}
    X = X^0 + X^1\,,
        \qquad%
    \overline{X} = X^0 - X^1\,.
\ee
Integrating out $\lambda$ and $\bar{\lambda}$ in \eqref{eq:lla} gives back the original action \eqref{eq:hs}.
The zero $\alpha'$ limit leads to the low-energy sigma model that defines nonrelativistic string theory~\cite{Gomis:2000bd},
\be \label{eq:seff}
    S_\text{eff} = \frac{1}{4\pi\alpha'_\text{eff}} \int_\Sigma d^2 \sigma \lr \p_\alpha X^{A'} \, \p^\alpha X^{A'} + \lambda \, \bar{\p} X + \bar{\lambda} \, \p X \rr.
\ee
The one-form fields $\lambda$ and $\bar{\lambda}$ are Lagrange multipliers that impose the (anti-)holomorphic constraints $\bar{\p} X = \p \overline{X} = 0$\,. The worldsheet theory \eqref{eq:seff} is invariant under the global symmetry,
\be \label{eq:gbx}
    \delta_\text{G} X^A = 0\,, 
        \qquad
    \delta_\text{G} X^{A'} = \Lambda^{A'}{}_A \, X^A,
\ee
supplemented with appropriate transformations acting on $\lambda$ and $\bar{\lambda}$\,. This symmetry acting on worldsheet fields corresponds to a stringy generalization of the Galilei boost in the target space. The target space therefore enjoys a nonrelativistic isometry group. We already required that $X^1$ be compactified over a circle of radius $R$\,, which is essential for the nonrelativistic closed string spectrum to be non-empty~\cite{Klebanov:2000pp, Gomis:2000bd, Danielsson:2000gi}. This can be seen by applying the $\alpha' \rightarrow 0$ limit to the dispersion relation for relativistic closed strings in the background field configuration from \eqref{eq:hgb2}. The relativistic dispersion relation is
\be \label{eq:rdr}
    \lr \varepsilon + \frac{w R}{\alpha'} \rr^2 - \frac{\alpha'_\text{eff}}{\alpha'} \, k_{A'} k_{A'} = \frac{n^2}{R^2} + \frac{w^2 R^2}{\alpha'{}^2} + \frac{2}{\alpha'} \lr N + \tilde{N} - 2 \rr,
\ee
together with the level-matching condition $n w = \tilde{N} - N$\,.
Here, $\varepsilon$ is the closed string energy in absence of the $B$-field. Recall that $w$ is the winding number in $X^1$, $n$ is the KK number, $k_{A'}$ is the transverse momentum, and $N$ and $\tilde{N}$ are string excitation numbers. 
In the $\alpha' \rightarrow 0$ limit, we find that \eqref{eq:rdr} becomes
\be \label{eq:drcs}
    \varepsilon = \frac{\alpha'_\text{eff}}{2 w R} \, \ls k_{A'} k_{A'} + \frac{2}{\alpha'_\text{eff}} \lr N + \tilde{N} - 2 \rr \rs,
\ee
which is the dispersion relation in nonrelativistic closed string theory. The level-matching condition is unchanged under this limit. This dispersion relation is only well-defined for states with nonzero windings.
Note that the KK number $n$ in $X^1$ only shows up in the level-matching condition but not the dispersion relation explicitly. 

A closed string amplitude in nonrelativistic string theory also factorizes into a quadratic product of open string amplitudes. This nonrelativistic analog of the KLT relation can be obtained by taking a zero $\alpha'$ limit of the relativistic KLT relation for winding string amplitudes in the presence of a critical $B$-field. We therefore start with the KLT relation \eqref{eq:rcsamp} that is supplemented with the kinematic data in \eqref{eq:cskbf}, 
but now with $B = -1$ to match with the $B$-field configuration in \eqref{eq:hgb2}. 
Moreover, the transverse momentum $k_{A'}$ in \eqref{eq:cskbf} is rescaled by a factor $\sqrt{\alpha'_\text{eff}/\alpha'}$\,, which takes into account the metric configuration in \eqref{eq:hgb2}.
These lead to a new set of kinematic data,
\begin{subequations} \label{eq:nklr}
\begin{align}
    K_{\text{L}} & = \lr \varepsilon + \frac{w R}{\alpha'} \,, \, \frac{n}{R} - \frac{wR}{{\alpha}'}\,, \, \sqrt{\frac{\alpha'_\text{eff}}{\alpha'}} \, k_{A'} \rr, \\[2pt]
    K_{\text{R}} & = \lr \varepsilon + \frac{w R}{\alpha'}\,, \, \frac{n}{R} + \frac{wR}{{\alpha}{}'}\,, \, \sqrt{\frac{\alpha'_\text{eff}}{\alpha'}} \, k_{A'} \rr.
\end{align}
\end{subequations}
Using \eqref{eq:nklr}, we find that the kinematic variable in \eqref{eq:kinv} is modified to be 
\begin{align}
\begin{split}
    \alpha' K_{\text{L}i} \cdot K_{\text{L}j} & = - \alpha' \lr \varepsilon_i + \frac{w_i R}{\alpha'} \rr \, \lr \varepsilon_j + \frac{w_j R}{\alpha'} \rr \\[2pt]
    & \quad + \alpha' \lr \frac{n_i}{R} - \frac{w_i R}{\alpha'} \rr \! \lr \frac{n_j}{R} - \frac{w_j R}{\alpha'} \rr + \alpha'_\text{eff} \, k_i^{A'} k_j^{A'},
\end{split}
\end{align}
which, unlike \eqref{eq:daklkl}, is now non-singular under the limit $\alpha' \rightarrow 0$\,, with
\begin{subequations} \label{eq:zaklkr}
\begin{align}
    \alpha' K_{\text{L}i} \cdot K_{\text{L}j} \rightarrow - \lr w_i \, \varepsilon_j + w_j \, \varepsilon_i \rr R + \alpha'_\text{eff} \, k_i^{A'} k^{A'}_j - \lr n_i \, w_j + n_j \, w_i \rr,
\end{align}
Similarly, we have in the $\alpha' \rightarrow 0$ limit,
\begin{align}
    \alpha' K_{\text{R}i} \cdot K_{\text{R}j} \rightarrow 
    - \lr w_i \, \varepsilon_j + w_j \, \varepsilon_i \rr R + \alpha'_\text{eff} \, k_i^{A'} k^{A'}_j + \lr n_i \, w_j + n_j \, w_i \rr,
\end{align}
\end{subequations}
where the sign of each winding number is flipped.
Therefore, the KLT factorization of a nonrelativistic closed string amplitude is in form the same as \eqref{eq:rcsamp}, except that the kinematic variables $\alpha' K_{\text{L}i} \cdot K_{\text{L}j}$ and $\alpha' K_{\text{R}i} \cdot K_{\text{R}j}$ are replaced with the ones in \eqref{eq:zaklkr}.

We already learned in \S\ref{sec:rklt} that, in the KLT relation for winding string amplitudes, the open string amplitudes involve open strings that end on an array of D-branes, which are equidistantly localized along the compactified circle. The separation between each consecutive D-branes is $L = 2 \pi \alpha' / R$\,, which is T-dual to the circumference of the compactified circle. In the $\alpha' \rightarrow 0$ limit, this distance $L$ becomes zero, giving rise to a stack of $n_+ + 1$ coinciding D-branes, where $n_+$ is the total momentum number of the incoming closed strings (see Fig.~\ref{fig:fpcsa}). 

We also learned from \S\ref{sec:wkltbf} that, in the presence of a $B$-field, the relativistic KLT relation involves amplitudes with open strings ending on D-branes carrying electric potentials. Namely, the $s$-th D-brane carries an electric potential $V_s = s B / R$ as in \eqref{eq:vsfinal}. 
To make contact with nonrelativsitic string theory, we set $B = -1$ followed by sending $\alpha'$ to zero. However, since $V_s$ is independent of $\alpha'$\,, it remains unchanged under the limit. As we have seen before, the $U(n_++1)$ gauge group on the stack of $n_++1$ D-branes is broken into its Cartan subalgebra $U(1)^{n_+ + 1}$, with the gauge potential $A_0 = \text{diag} \bigl( V_1, \cdots, V_{n_++1} \bigr)$\,.

To summarize, the KLT relations in nonrelativistic string theory is in form the same as \eqref{eq:rcsamp}, but with the kinematic variables $\alpha' K_{\text{L}i} \cdot K_{\text{L}j}$ and $\alpha' K_{\text{R}i} \cdot K_{\text{R}j}$ replaced with the expressions in \eqref{eq:zaklkr}. Moreover, the open string amplitudes involved in the KLT relation consist of open strings that end on a stack of coinciding $n_+ + 1$ D-branes, with each D-brane carrying an electric potential such that there is a constant electric field between the D-branes. In the next section, we will derive the same KLT relations in nonrelativistic string theory from first principles, without resorting to the zero $\alpha'$ limit.

\section{KLT Factorizations in Nonrelativistic String Theory} \label{sec:kltfnsa}

We now switch our attention away from relativistic string theory and give a brief review of nonrelativistic string theory. We will take the defining action \eqref{eq:seff} for nonrelativistic string theory as the starting point, without relying on the low energy limit discussed in \S\ref{sec:zalwkltr}. 

\subsection{Nonrelativistic Closed String Theory}\label{enst}

We first focus on the closed string sector. As preluded in \S\ref{sec:zalwkltr}, nonrelativistic string theory in flat spacetime is defined by the action principle \eqref{eq:seff}, \emph{i.e.},
\be \label{eq:nrsa}
    S = \frac{1}{4\pi\alpha'_\text{eff}} \int_\Sigma d^2 \sigma \left( \p_\alpha X^{A'} \, \p^\alpha X^{A'} + \lambda \, \bar{\p} X + \bar{\lambda} \, \p \overline{X} \right),
\ee
We briefly summarize the definitions that we already gave earlier in \S\ref{sec:kltfwsdb} for the action \eqref{eq:nrsa} below. Here, $\sigma^\alpha = (\tau, \sigma)$ are coordinates on the relativistic worldsheet, with $\tau$ the imaginary time. The worldsheet $\Sigma$ is mapped to a foliated spacetime manifold $\CM$ by the worldsheet coordinates $X^\mu = (X^A,X^{A'})$, with $A = 0,1$ and $A' = 2, \dots, d-1$\,. The coordinates $X^A$ and $X^{A'}$ define the longitudinal and transverse sectors, respectively. 
The derivatives $\p$ and $\bar{\p}$ are defined in \eqref{eq:pbp} and the lightlike coordinates $X$ and $\overline{X}$ are defined in \eqref{eq:XbX}. The one-form fields $\lambda$ and $\bar{\lambda}$ are Lagrange multipliers that impose the~(anti-)holomorphic constraints $\bar{\p} X = \p \overline{X} = 0$\,. The target space enjoys a nonrelativistic global isometry group, including the string Galilei boost symmetry in \eqref{eq:gbx} that relates the transverse and longitudinal sectors. 
We also recall that the compactification of $X^1$ over a circle of radius $R$ is necessary for the closed string spectrum to be non-empty. 
The closed string spectrum enjoys a Galilean invariant dispersion relation \eqref{eq:drcs}, which we transcribe below:
\be \label{eq:drcs2}
    \varepsilon = \frac{\alpha'_\text{eff}}{2 w R} \, \ls k_{A'} k_{A'} + \frac{2}{\alpha'_\text{eff}} \lr N + \tilde{N} - 2 \rr \rs,
\ee
where $\varepsilon$ is the energy, $w$ is the winding number in $X^1$, $k_{A'}$ is the transverse dispersion relation, and $N$ and $\tilde{N}$ are string excitation numbers. The KK number $n$ in $X^1$ enters the dispersion relation via the level-matching condition $n \, w = \tilde{N} - N$\,.

In radial quantization, we use the conformal mapping, $z = e^{\tau + i \sigma}$ and $\bar{z} = e^{\tau - i \sigma}$, in terms of which the string action \eqref{eq:nrsa} becomes
\be\label{nrsacomplex}
    S = \frac{1}{4\pi\alpha'_\text{eff}} \int_\mathbb{C} d^2 z \lr 2 \, \p_z X^{A'} \, \p_{\bar{z}} X^{A'} + \lambda_z \, \p_{\bar{z}} X + \lambda_{\bar{z}} \, \p_z \overline{X} \rr,
\ee
where 
\begin{subequations} \label{eq:llpp}
\begin{align}
    \lambda_z & = - i \lambda / z\,, 
        &%
    \p_z & = \frac{1}{2}(\p_\tau-i \p_\sigma)\,, \\[2pt]
    \lambda_{\bar{z}} & = i \bar{\lambda} / \bar{z}\,, 
        &%
    \p_{\overline{z}} & = \frac{1}{2}(\p_\tau+i\p_\sigma).
\end{align}
\end{subequations}
The operator product expansions (OPEs) between different worldsheet fields are
\begin{subequations} \label{eq:opes}
\begin{align}
    & :\!\lambda_z (z_1) \, X(z_2) \!: \, \sim - \frac{2 \, \alpha'_\text{eff}}{z_1 - z_2}\,,
        \qquad
    :\! \lambda_{\bar{z}} (\bar{z}_1) \, \overline{X}(\bar{z}_2) \!: \, \sim - \frac{2 \, \alpha'_\text{eff}}{\bar{z}_{1} - \bar{z}_2}\,, \\[2pt]
    & : \! X^{A'} (z_1\,, \bar{z}_1) \, X^{B'} (z_2\,, \bar{z}_2) \!: \, \sim - \frac{\alpha'_\text{eff}}{2} \, \delta^{A'B'} \, \ln |z_{1} - z_2|^2.
\end{align}
\end{subequations}
Following \cite{Yan:2021lbe}, these OPEs can be written in a compact way as follows. First, we introduce a local redefinition of the one-form fields $\lambda$ and $\bar{\lambda}$\,,
\be \label{eq:frd}
    \lambda_z = - 2 \, \p_z X'\,,
        \qquad
    \lambda_{\bar{z}} = 2 \, \p_{\bar{z}} \overline{X}{}^\prime\,.
\ee
The auxiliary coordinates $X' = X'(z)$ and $\overline{X}{}' = \overline{X}{}'(\bar{z})$ are T-dual to $X$ and $\overline{X}$\,, respectively. The field redefinitions \eqref{eq:frd} involve time derivatives and need to be treated with care in the path integral. Nevertheless, it is valid to perform a direct substitution of \eqref{eq:frd} in the operator formalism. We further define
\begin{subequations} \label{eq:rdfp}
\begin{align}
	\varphi^0_\text{L} (z) & = \tfrac{1}{2} (X+X')\,, 
	    \qquad%
	\varphi_\text{R}^0 (\bar{z}) = \tfrac{1}{2} (\overline{X} - \overline{X}{}^\prime)\,, \\[2pt]
	\varphi^1_\text{L} (z) & = \tfrac{1}{2} (X-X')\,, 
	    \qquad%
	\varphi_\text{R}^1 (\bar{z}) = \tfrac{1}{2} (\overline{X} + \overline{X}{}^\prime)\,,
\end{align}
\end{subequations}
together with $X^{A'} = \varphi^{A'} (z) + \bar{\varphi}^{A'} (\bar{z})$\,. 
In terms of the new variables $\varphi^\mu$, the OPEs in \eqref{eq:opes} become
\be
    \varphi^\mu_\text{L} (z_1) \, \varphi^{\nu}_\text{L} (z_2) \sim - \frac{\alpha'_\text{eff}}{2} \, \eta^{\mu\nu} \, \ln \bigl( z_1- z_2 \bigr)\,,
        \qquad%
    \varphi_\text{R}^\mu(\bar{z}_1) \, \varphi_\text{R}^{\nu} (\bar{z}_2) \sim - \frac{\alpha'_\text{eff}}{2} \, \eta^{\mu\nu} \, \ln \bigl( \bar{z}_1 - \bar{z}_2 \bigr)\,,
\ee
which take the same form of the OPEs between the worldsheet fields that play the role of the target-space coordinates in relativistic string theory. These OPEs coincide with the ones from the string action for relativistic string theory,
\be \label{eq:sphi}
    S_\varphi = \frac{1}{4\pi\alpha'_\text{eff}} \int d^2 \sigma \, \p_\alpha \varphi^\mu \, \p^\alpha \varphi_\mu\,,
\ee
where $\varphi^\mu = \varphi^\mu_\text{L} + \varphi^\mu_\text{R}$\,. With appropriate spacetime compactifications, the closed string tachyon vertex operator takes the same form as in \eqref{eq:cv}, with the left- and right-moving momentum $K_\text{L}$ and $K_\text{R}$\,, respectively.  
In practice, when the OPEs are concerned, physical quantities in nonrelativistic string theory can be obtained from relativistic string theory by plugging in the mappings in \eqref{eq:rdfp}. 

The vertex operators can be constructed using the variables $\varphi^\mu_\text{L,\,R}$ as well. 
We start with a detour to first identify different quantum numbers. Define the quantum numbers $q$ and $\bar{q}$ that are respectively conjugate to $X'$ and $\overline{X}{}'$\,. These quantum numbers are eigenvalues of the operators,
\be \label{eq:qqb}
    \hat{q} = - \frac{1}{2\pi \alpha'_\text{eff}} \oint_\CC dz \, \p_z X\,,
        \qquad%
    \hat{\bar{q}} = - \frac{1}{2\pi \alpha'_\text{eff}} \oint_\CC d\bar{z} \, \p_{\bar{z}} \overline{X}\,,
\ee
where $\CC$ is a contour that encloses the vertex operator counterclockwise. In our case, only $X^1$ is compactified, and $q = - \bar{q} = w R / \alpha'_\text{eff}$. We also define the quantum numbers $p$ and $\bar{p}$ that are respectively conjugate to $X$ and $\overline{X}$\,. These quantum numbers are eigenvalues of the momentum operators
\be
    \hat{p} = - \frac{1}{2\pi \alpha'_\text{eff}} \oint_\CC dz \, \p_z X'\,,
        \qquad%
    \hat{\bar{p}} = - \frac{1}{2\pi \alpha'_\text{eff}} \oint_\CC d\bar{z} \, \p_{\bar{z}} \overline{X}'\,.
\ee
Writing $p = \frac{1}{2} (p^{}_0 + p^{}_1)$ and $\bar{p} = \frac{1}{2} (p^{}_0 - p^{}_1)$\,, we observe that $p^{}_A$ is the longitudinal momentum. Note that $p^{}_0 = \varepsilon$ is the energy. As $X^1$ is compactified, $p_1 = n / R$ is quantized, with $n \in \mathbb{Z}$ the KK number. Therefore, the closed string tachyon vertex operator is
\be \label{eq:cstvo}
    \CV = \exp \bigl( i \pi \, n \, \hat{w} \bigr) \, : \! \exp \bigl( i \, K_{\!A'} \, X^{\!A'} + i \, p^{}_{\!A} X^{\!A} + i \, q^A X'_{\!A} \bigr) \! :.
\ee
We defined $X'_0 = \frac{1}{2} \lr X' + \overline{X}{}' \rr$ and $X'_1 = \frac{1}{2} \lr X' - \overline{X}{}' \rr$. Moreover, $q^0 = q + \bar{q}$ and $q^1 = q - \bar{q}$\,. The phase factor $e^{i \pi \, n \, \hat{w}}$ is a cocycle factor that is required for the vertex operators to commute. 
In terms of $\varphi_\text{L,\,R}^\mu$ in \eqref{eq:rdfp}, the vertex operator \eqref{eq:cstvo} becomes
\be \label{eq:cv}
    \CV = \exp \ls \tfrac{i}{4} \, \pi \alpha'_\text{eff} \bigl( K_\text{L} - K_\text{R} \bigr) \cdot \bigl( \hat{K}_\text{L} + \hat{K}_\text{R} \bigr) \rs \, \exp \bigl( i K^{}_{\text{L}} \cdot \varphi_\text{L} + i K^{}_{\text{R}} \cdot \varphi_\text{R} \bigr)\,,
\ee
where
\be \label{eq:qn}
    K^{}_{\text{L}\,\mu} = ( p + q\,, \, p - q\,, \, k_{A'} )\,,
        \qquad%
    K^{}_{ \text{R}\, \mu } = ( \bar{p} - \bar{q}\,, \, \bar{p} + \bar{q}\,, \, k_{A'} )\,.
\ee
Higher-order vertex operators are constructed from dressing \eqref{eq:cv} up with derivatives acting on $\varphi_{\text{L,\,R}}^\mu$\,. For later use, we also note that it is possible to define the vertex operator as
\be 
    \CV = \exp \ls \tfrac{i}{4} \, \pi \alpha'_\text{eff} \, \bigl( K_\text{L} + K_\text{R} \bigr) \cdot \bigl( \hat{K}_\text{L} - \hat{K}_\text{R} \bigr) \rs \, \exp \bigl( i K^{}_{\text{L}} \cdot \varphi_\text{L} + i K^{}_{\text{R}} \cdot \varphi_\text{R} \bigr)\,,
\ee
where we chose to remove the branch cuts from permuting vertex operators differently. This change in the definition of vertex operators only contributes an overall sign in front of the amplitude and thus not affecting the associated cross section. 
\subsection{Nonrelativistic Open Strings} \label{sec:nos}

Now, we consider in nonrelativistic string theory a D-brane that is transverse to the longitudinal spatial direction $X^1$\,, which is compactified. Open strings ending on such a D-brane satisfy the Dirichlet boundary condition,
\be
    \delta X^1 \big|_{\p \Sigma} = 0\,.    
\ee
At tree level, we require that the worldsheet $\Sigma$ is a strip with $\tau \in \mathbb{R}$ and the boundary $\p\Sigma$ resides at $\sigma = 0\,, \pi$\,. For simplicity, we require that all the other directions satisfy the Neumann boundary conditions, with $\p_\sigma X^0 \big|_{\p\Sigma} = \p_\sigma X^{A'} \big|_{\p\Sigma} = 0$\,. Varying the action \eqref{eq:nrsa} with respect to $X^\mu$ further requires
\be
    \lambda + \bar{\lambda} \, \big|_{\p\Sigma} = 0\,.
\ee
The (anti-)holomophic conditions $\bar{\p} X = \p \overline{X} = 0$ induce an extra boundary condition,
\be
    \p_\sigma X^1 - i \, \p_\tau X^0 \, \big|_{\p\Sigma} = 0\,.
\ee
This theory has a nonrelativistic open string spectrum, with
\be \label{eq:nrosdr}
    \varepsilon = \frac{\alpha'_\text{eff}}{2 w R} \ls k_{A'} k_{A'} + \frac{1}{\alpha'_\text{eff}} \lr N - 1 \rr \rs.
\ee
Here, the winding number $w$ describes how many times an open string wraps around $X^1$, with both the ends of the string anchored on the same D-brane. In the case where the open string ends on different D-branes, $w$ can be fractional.

In terms of the complex variables $z = e^{\tau + i \sigma}$ and $\bar{z} = e^{\tau - i \sigma}$\,, the strip is mapped to be the upper half of the complex plane, with the boundary being the real axis at $z = \bar{z}$\,. Using the definitions from \eqref{eq:llpp}, the boundary conditions on $X^A$ become \cite{Danielsson:2000mu}
\be
    \p_z X - \p_{\bar{z}} \overline{X} \big|_{z = \bar{z}} = 0\,, 
        \qquad%
    \lambda_z - \lambda_{\bar{z}} \big|_{z=\bar{z}} = 0\,,
\ee
supplemented with the (anti-)holomorphic conditions $\p_{\bar{z}} X \big|_{z=\bar{z}} = \p_z \overline{X} \big|_{z=\bar{z}} = 0$\,.  
Using the local field redefinitions in \eqref{eq:frd}, we find
\be
    \p_z X' + \p_{\bar{z}} \overline{X}{}' \big|_{z=\bar{z}} = 0\,,
\ee
also supplemented with the (anti-)holomorphic conditions $\p_{\bar{z}} X' \big|_{z=\bar{z}} = \p_z \overline{X}{}' \big|_{z=\bar{z}} = 0$\,. Finally, also taking into account the Neumann boundary conditions $\p_\sigma X^{A'} \big|_{\p\Sigma} = 0$\,, and in terms of the variables $\varphi_{\text{L,\,R}}^\mu$ in \eqref{eq:rdfp}, we find that the above boundary conditions become
\be \label{eq:bcphi}
      \p_z \varphi^{\mu}_\text{L} (z) - \p_{\bar{z}} \varphi^{\mu}_\text{R} (\bar{z}) \big|_{z=\bar{z}} = 0\,,
\ee
which imply that $\p_\sigma \varphi^\mu \big|_{\p\Sigma} = 0$\,,
with $\varphi^\mu = \varphi^\mu_\text{L} + \varphi^\mu_\text{R}$\,.
Therefore, in terms of the variable $\varphi^\mu$ that mixes $X^A$ and ${X'}{}^A$, all $\varphi^\mu$ directions effectively satisfy Neumann boundary conditions. We emphasize that, even though various expressions in terms of $\varphi^\mu$ are formally the same as the associated ones in relativistic string theory, physically, this is only the consequence of a redefinition of the target-space coordinates and their duals in nonrelativistic string theory.

The OPEs that satisfy the Dirichlet boundary condition in $X^1$ and Neumann boundary conditions in the remaining directions are
\begin{subequations} 
\begin{align}
    & :\!\lambda_z (z_1) \, X(z_2) \!: \, \sim - \frac{2\alpha'_\text{eff}}{z_1 - z_2}\,,
        \qquad
    :\! \lambda_{\bar{z}} (\bar{z}_1) \, \overline{X}(\bar{z}_2) \!: \, \sim - \frac{2\alpha'_\text{eff}}{\bar{z}_1 - \bar{z}_2}\,, \\[2pt]
    & : \! X^{A'} \! (z_1\,, \bar{z}_1) \, X^{B'} \! (z_2\,, \bar{z}_2) \!: \, \sim - \frac{\alpha'_\text{eff}}{2} \, \delta^{A'B'} \lr \ln |z_1 - z_2|^2 + \ln |z_1 - \bar{z}_2|^2 \rr.
\end{align}
\end{subequations}
In terms of $\varphi^\mu_\text{L,\,R}$\,, the open string OPEs on the worldsheet boundary $\p\Sigma$ are obtained by restricting the following expressions to $z = \bar{z} = y$ on $\p\Sigma$\,, with $y \in \mathbb{R}$\,: 
\be \label{eq:operelphi}
    : \! \varphi^\mu \! (z_1\,, \bar{z}_1) \, \varphi^\nu \! (z_2\,, \bar{z}_2) \!: \, \sim - \frac{\alpha'_\text{eff}}{2} \, \eta^{\mu\nu} \lr \ln |z_1 - z_2|^2 + \ln |z_1 - \bar{z}_2|^2 \rr.
\ee
The same boundary OPEs can be reproduced by using the action \eqref{eq:sphi} with the Neumann boundary condition $\p_\sigma \varphi^\mu \big|_{\p\Sigma} = 0$\,.  

The open string vertex operators can be constructed by restricting the closed string vertex operators \S\ref{enst} to be on the boundary. For a open string tachyons state, with the ends of the open string anchored on two D-branes located in the compactified direction $X^1$ and separated by a distance $L$\,, the associated vertex operator is
\be \label{eq:vow}
    \CV_\text{open} = \, :\!\exp \bigl[ \, i \, K_\text{L} \cdot \varphi^{}_\text{L} (y) + i \, K_\text{R} \cdot \varphi^{}_\text{R} (y) \, \bigr] \!:\,,
\ee
where $K_{\text{L,\,R}\mu}$ can be read off from \eqref{eq:qn} by setting $p^{}
_1 = 0$ and $q = - \bar{q} = W R / \alpha'_\text{eff}$\,, with
\be \label{eq:klrW}
    K_{\text{L}\mu} = K_{\text{R}\mu} = K_\mu \equiv \lr \frac{\varepsilon}{2} + \frac{W R}{\alpha'_\text{eff}}\,,\, \frac{\varepsilon}{2} - \frac{W R}{\alpha'_\text{eff}}\,, k_{A'} \rr,
        \qquad%
    W = w + \frac{L}{2\pi R}\,.
\ee
Here, $\varepsilon = p^{}_0$ is the energy and $W$ is the total winding number. We defined $w$ to be the integer part that counts how many times the open string wraps around the compactified circle in $X^1$\,. Note that the momentum $p^{}_1$ in the compactified direction is set to zero since an open string with Dirichlet boundary conditions on both ends cannot carry any collective momentum. As a result, the vertex operator \eqref{eq:vow} reduces to
\be \label{eq:vowsim}
    \CV_\text{open} = g_{\text{0}}\, :\! \exp \! \big[ i \, K \cdot \varphi (y) \bigr] \!:\,,
\ee
where $\varphi = \varphi^{}_\text{L} + \varphi^{}_\text{R}$ satisfies the OPE
\be \label{eq:opeff}
    :\! \varphi^\mu (y_1) \, \varphi^\nu (y_2) \!: \, \sim - 2 \, \alpha'_\text{eff} \, \eta^{\mu\nu} \, \ln \bigl( y_1 - y_2
    \bigr)\,.  
\ee
The string amplitude then takes the same form as in relativistic string theory, except that the kinematic data is given by \eqref{eq:klrW}.

\subsection{KLT Factorization of Nonrelativistic Closed String Amplitudes} \label{sec:kltfncsa}

The above rewriting of the OPEs and vertex operators in terms of $\varphi^\mu_{\text{L,\,R}}$ is particularly useful for evaluating string amplitudes in nonrelativistic string theory: we simply need to take well-known results of relativistic string amplitudes and plug in the kinematic data \eqref{eq:qn}! For example, the $\CN$-point closed string amplitude takes the Virasoro-Shapiro form,
\begin{align} \label{eq:mcn}
    \tilde{\CM}^\CN_\text{c} = \tilde{C} (1, \dots, \, \CN) \, \int_{\mathbb{C}^{\CN-3}} d^2 z_2 \cdots d^2 z_{\CN-2}  \, \prod_{\substack{i,j = 1\\i<j}}^{\CN-1} z_{ji}^{\frac{1}{2} \alpha'_\text{eff} K_{\text{L}i} \cdot K_{\text{L}j}} \, \bar{z}_{ji}^{\frac{1}{2} \alpha'_\text{eff} K_{\text{R}i} \cdot K_{\text{R}j}} .
\end{align}
We have applied the M\"{o}bius transformation to fix $z_1 = 0$\,, $z_{\CN-1} = 1$ and $z_\CN = \infty$\,. The factor $\tilde{C} (1, \dots, \, \CN)$ comes from the cocycle factors in vertex operators, and takes the following form:
\be \label{eq:newcf}
    \tilde{C} (i_1, \dots, \, i_\CN) = \exp \lr i \pi {\sum}_{\substack{p,\,q=1\\p<q}}^\CN w_{i_p} n_{i_q} \rr\!.
\ee
Using the mapping in \eqref{eq:qn}, we find
\begin{subequations} \label{eq:kllrr}
\begin{align}
    \alpha'_\text{eff} \, K_{\text{L}i} \cdot K_{\text{L}j} = - \lr w_i \, \varepsilon_j + w_j \, \varepsilon_i \rr R + \alpha'_\text{eff} \, k_i^{A'} k^{A'}_j - \lr n_i \, w_j + n_j \, w_i \rr, \label{eq:klkl} \\[2pt]
    \alpha'_\text{eff} \, K_{\text{R}i} \cdot K_{\text{R}j} = - \lr w_i \, \varepsilon_j + w_j \, \varepsilon_i \rr R + \alpha'_\text{eff} \, k_i^{A'} k^{A'}_j + \lr n_i \, w_j + n_j \, w_i \rr. \label{eq:krkr}
\end{align}
\end{subequations}
Note that the above expressions agree with \eqref{eq:zaklkr}, which we obtained by taking a zero Regge slope limit in relativistic string theory.  We will return to this point later after examining the open string amplitudes.

We apply the same procedure as we did for nonrelativistic closed strings to obtain amplitudes in nonrelativistic open string theory.  In the simplest example, we consider amplitudes for scattering of nonrelativistic open strings, which end on a single D-brane located at $x^1$ in the compactified $X^1$ direction. Such open strings satisfying the Dirichlet boundary condition in $X^1$ carry nonzero winding numbers that describe how many times the open string wraps around $X^1$, with the ends anchored on D-branes that are transverse to $X^1$. However, the momentum along $X^1$ is zero, required by the Dirichlet boundary condition.
According to \S\ref{enst}, we can equivalently calculate relativistic open string amplitudes with Neumann boundary conditions in all directions, but with the kinematic data in \eqref{eq:qn}, which now becomes
\be \label{eq:Kmu}
    K_\text{L,\,R}^M = K^M = \lr \frac{\varepsilon}{2} + \frac{w R}{\alpha'_\text{eff}}\,, \, \frac{\varepsilon}{2} - \frac{w R}{\alpha'_\text{eff}}\,, \, k_{A'} \rr.
\ee
Here, $w$ is fractional if the open string ends on two D-branes that are separated by a distance $L$ along $X^1$. The $\CN$-point open string amplitude is 
\be \label{eq:mon}
    \tilde{\CM}_\text{o} (1, \dots, \CN) = \int_{0 < y_2 < \cdots < y_{\CN-1} < \infty} dy_2 \, \cdots \, dy_{\CN-2}  \, \prod_{\substack{i,j = 1\\i<j}}^{\CN-1} |y_{ij}|^{2 \alpha'_\text{eff} K_{i} \cdot K_j},
\ee
where $y_{ij} = y_i - y_j$\,. 
We have used the SL(2,\,$\mathbb{R}$) symmetry to fix $y_0 = 1\,, y_{\CN-1} = 1$ and $y_\CN = \infty$\,.  The kinematic data $K^M_i$ for the $i$-th string is defined in \eqref{eq:Kmu}. 

The $\CN$-point KLT relation is given by 
the following factorization of the closed string amplitude \eqref{eq:mcn}:
\begin{align} \label{eq:nrklt}
\begin{split}
    \tilde{\CM}_\text{c}^\CN & = (-1)^{\CN-3} \sum_{\rho,\,\sigma} \tilde{C}(1, \sigma, \, \CN-1, \, \CN) \, \mathcal{S}_\text{L} [\rho|\sigma]^{}_{K_{\text{L}1}} \\[-4pt]
    & \hspace{3cm} \times \tilde{\CM}_\text{L} \lr \rho, \,1, \, \CN-1, \, \CN \rr \, \tilde{\CM}_\text{R} \lr 1, \, \sigma, \, \CN-1, \, \CN \rr,
\end{split}
\end{align}
where $\CM_\text{c}^\CN$ is given in \eqref{eq:mcn} and $\CM_\text{L,\,R}$ is essentially given by \eqref{eq:mon}, but with the rescaling $\alpha'_\text{eff} \rightarrow \frac{1}{4} \, \alpha'_\text{eff}$ and $R \rightarrow \frac{1}{4} \, R$\,, such that
\be \label{nrosamp}
    \tilde{\CM}_\text{L,\,R} (1, \dots, \CN) = \int_{0 < y_2 < \cdots < y_{\CN-1} < 1} dy_2 \, \cdots \, dy_{\CN-2}\, \prod_{\substack{i,j = 1\\i<j}}^{\CN-1} |y_{i} - y_{j}|^{\frac{1}{2} \, \alpha'_\text{eff} \, K_{\text{L,\,R}i} \cdot K_{\text{L,\,R}j}}\,,
\ee
and the momentum kernel is the same as in \eqref{momentumkernel}. This KLT relation is in form the same as the standard ones, except that $K_{\text{L,\,R}i}$ have the form of \eqref{eq:Kmu}. We start with the kinematic quantities $K_{\text{L,R}i}$ from the closed string side, as defined in \eqref{eq:qn}, with
\begin{subequations}\label{eq:closedMomenta}
\begin{align}
    K^{M}_{\text{L}} & = \Bigl( \frac{\varepsilon}{2} + \frac{n}{2R} + \frac{w R}{\alpha'_\text{eff}}\,, \, \frac{\varepsilon}{2} + \frac{n}{2R} - \frac{w R}{\alpha'_\text{eff}}\,, \, k_{A'} \Bigr)\,, \label{eq:KL} \\[2pt]
    K^{M}_{\text{R}} & = \Bigl( \frac{\varepsilon}{2} - \frac{n}{2R} + \frac{w R}{\alpha'_\text{eff}}\,, \, \frac{\varepsilon}{2} - \frac{n}{2R} - \frac{w R}{\alpha'_\text{eff}}\,, \, k_{A'} \Bigr)\,, \label{eq:KR}
\end{align}
\end{subequations}
where we suppressed the subscript $i$ that labels different string states.
To answer which open string amplitudes \eqref{nrosamp} are associated with the expressions in \eqref{nrosamp}, we focus on $\CM_\text{L}$ therein for concreteness. We consider nonrelativistic open strings that end on D-branes located in $X^1$. Then, in the KLT relation, the winding number $w_i$ of the $i$-th closed string is mapped to the winding number of the $i$-th open string that ends on the same D-brane after wrapping around $X^1$ circle for $w_i$ times. Such open strings have zero momentum in $X^1$. We thus have $p = \bar{p} = \varepsilon / 2$ and $q = - \bar{q} = wR / \alpha'_\text{eff}$\,.
The associated kinematic data $K^M$ is the  same as in \eqref{eq:Kmu}. To identify this $K^M$ for an open string state with $K^M_\text{L}$ in \eqref{eq:KL}, we need to further shift the energy by $n / R$\,. This is achieved by introducing a stack of D-branes that are transverse to $X^1$ and located at the same point in $X^1$. We also have to assign to the $s$-th D-brane an electric potential $V_s = - s L / (2\pi\alpha'_\text{eff})$\,, where $L = 2 \pi \tilde{R}$\,, with $\tilde{R}$  the T-dual of the radius of the compact circle. Consider an open string with its two ends anchored on the $s$-th and $(s+n)$-th D-brane. In this case, we find the following shift in the energy:
\be \label{eq:ees}
    \varepsilon \rightarrow \varepsilon + \frac{n}{R}\,.
\ee
Upon rescaling $\alpha'_\text{eff} \rightarrow \alpha'_\text{eff}/4$ and $R \rightarrow R/4$\,, together with $L \rightarrow L / 4$ such that $V_s$ remains unchanged, the resulting amplitude that involves nonrelativistic open strings ending on a tack of electrically charged D-branes matches with $\CM_\text{L}$ in \eqref{nrosamp}. 
To get $\CM_\text{R}$\,, we just need to consider the D-brane configuration but with an opposite sign in the electric potential $V_s$\,. We furthermore see that the KLT relation in \eqref{eq:nrklt} is the same as the one derived in \S\ref{sec:zalwkltr} from taking a low energy limit of a KLT relation in relativistic string theory that involves a spacelike compactification along a constant $B$-field.~\footnote{It is also possible to construct KLT relations in nonrelativistic string theory in the presence of a $B$-field along the compact circle.
Since the KLT relation in nonrelativistic string theory requires a stack of D-branes that are located at the same point in the compact circle, introducing a $B$-field in $X^1$ shifts the energies of the closed and open strings by the same amount. Therefore, unlike the situation in relativistic string theory, no extra ingredients are required in the presence of a $B$-field now.} However, the cocycle contribution $\tilde{C}$ in \eqref{eq:newcf} that we now have is different from the one that we called $C$ in \eqref{eq:cfori}, with the latter being used in \S\ref{sec:zalwkltr}. This difference only introduces an overall sign in front of the full amplitude, because
\be \label{eq:ovasgn}
    C (i_1, \cdots, i_k) \, \tilde{C} (i_1, \cdots, i_k) = \exp \! \lr - i\pi \sum_{i=1}^k n_i \, w_i \rr
\ee
is independent of the ordering of vertex operators.

\subsection{String Theory in Discrete Light Cone Quantization} \label{sec:stdlcq}

It is not a coincidence that nonrelativistic string theory can be recast in the form of relativistic string theory when the OPEs in \eqref{eq:operelphi} are concerned. This hidden relativistic nature is made manifest by transforming the theory into a T-dual frame, where the discrete light cone quantization (DLCQ) of relativistic string theory arises. The DLCQ of string theory refers to relativistic string theory compactified on a lightlike circle. We now give an elementary review on how this lightlike compactification can be defined as a subtle infinite-boost limit \cite{Seiberg:1997ad} and its relation to nonrelativistic string theory \cite{Bergshoeff:2018yvt}. Via the T-duality, we will establish a KLT relation in the DLCQ of string theory, using the factorization of nonrelativistic closed string amplitudes considered in \S\ref{sec:kltfncsa}.

We start with the sigma model for relativistic string theory in flat spacetime,
\be \label{eq:srel}
    S_\text{rel.} = \frac{1}{4\pi\alpha'} \int d^2\sigma \, \p_\alpha Y{}^\mu \, \p^\alpha Y_\mu\,.
\ee
where the Euclidean worldsheet $\Sigma$ with coordinates $\sigma^\alpha = (\tau, \sigma)$ is mapped to the target space $\CM$ by a set of worldsheet fields $Y{}^\mu (\tau, \sigma)$, $\mu = 0, 1, \dots, d-1$\,.
We now compactify the spatial $
Y^1$ direction and impose the periodic boundary condition,
\be \label{eq:y1pbc}
    Y^1 \sim Y^1 + 2 \pi R_0\,.
\ee
Consider a large boost transformation at speed $v$ in $Y^1$,
\be \label{eq:xp}
    Y{}'{}^0 = \gamma \left( Y{}^0 + v \, Y{}^1 \right),
        \qquad
    Y{}'{}^1 = \gamma \left( Y^1 + v \, Y^0 \right),
\ee
where $\gamma = 1/ \sqrt{1 - v^2} \gg 1$\,. Define the lightlike coordinates 
$Y = Y^0 + Y^1$ and
$\overline{Y} = Y{}^0 - Y{}^1$.
Similarly, define $Y{}' = Y{}'{}^0 + Y{}'{}^1$ and $\overline{Y}{}' = {Y'}{}^0 - {Y'}{}^1$. Using \eqref{eq:xp}, we find
\be
    Y' = 2 \, \gamma \, Y + O(\gamma^{-1})\,,
        \qquad
    \overline{Y}{}' = \frac{\overline{Y}}{2 \, \gamma} + O(\gamma^{-3})\,.
\ee
The lightlike coordinates $Y'$ and $
\overline{Y}{}'$ satisfy the boundary conditions,
\be
    {Y'} \sim Y' + 4 \, R_0 \bigl[ \gamma \, \pi + O (\gamma^{-1}) \bigr]\,,
        \qquad%
    \overline{Y}{}' \sim \overline{Y}{}' - R_0 \left[ \frac{\pi}{2 \, \gamma} + O(\gamma^{-3}) \right]\,.
\ee
In the double-scaling limit $\gamma \rightarrow \infty$ and $R_0 \rightarrow 0$\,, while holding $\tilde{R} \equiv 2 \, \gamma R_0$ fixed, we find the periodic boundary condition
\be \label{eq:yp}
    Y' \sim Y' + 2 \pi \tilde{R}\,,
        \qquad
    \overline{Y}{}' \sim \overline{Y}{}',
\ee
i.e., $Y'$ is a lightlike direction that is compactified over a lightlike circle of radius $\tilde{R}$\,. This subtle limit is the typical procedure that is used to define the DLCQ of string theory in the literature. For example, see \cite{Seiberg:1997ad}.

It is instructive to apply the same double-scaling limit to the dispersion relation of closed strings. Before any limit is taken, the closed strings carry quantum numbers including the frequency $K_0$\,, the transverse momentum $K_{A'}$\,, with $A' = 2, \cdots, d-1$\,, and the string excitation numbers $N$ and $\tilde{N}$\,. Since we compactified $Y^1$ over a circle of radius $R_0$\,, we also have the KK number $\tilde{n}$ and winding number $\tilde{w}$\,. The dispersion relation before the double-scaling limit takes the following form:
\be \label{eq:drrel}
    K_0^2 - K_{A'} K_{A'} = \frac{\tilde{n}^2}{R_0^2} + \frac{\tilde{w}^2 R_0^2}{\alpha'{}^2} + \frac{2}{\alpha'} \left( N + \tilde{N} - 2 \right).
\ee
The finite-energy states in the double-scaling limit are determined by a large $\gamma$ expansion,
\be \label{eq:fes}
    K_0 - \frac{\tilde{n}}{R_0} = \frac{\varepsilon}{2 \, \gamma} + O(\gamma^{-2})\,.
\ee
where $\varepsilon$ is the effective energy of the physical state in the string spectrum.
Plugging \eqref{eq:fes} into the dispersion relation \eqref{eq:drrel}, we find
\be
    \frac{\varepsilon^2}{4 \, \gamma^2} + \frac{\tilde{n} \, \varepsilon}{\gamma \, R_0} - K_{A'} K_{A'} = \frac{\tilde{w}^2 R_0^2}{\alpha'{}^2} + \frac{2}{\alpha'} \left( N + \tilde{N} - 2 \right).
\ee 
In the double-scaling limit, we obtain the DLCQ dispersion relation
\be \label{eq:dlcqdr}
    \varepsilon = \frac{\tilde{R}}{2 \, \tilde{n}} \left[ K_{A'} K_{A'} + \frac{2}{\alpha'} \left( N + \tilde{N} - 2 \right) \right],
\ee
which enjoys a Galilei boost symmetry. 

Applying the double scaling limit to closed string amplitudes leads to DLCQ amplitudes, which are T-dual to amplitudes in nonrelativistic string theory.  To demonstrate this, we start with the $\CN$-point relativistic closed string amplitude that is in form the same as in \eqref{eq:mcn}, but with \eqref{eq:kllrr} replaced with
\begin{subequations} \label{eq:kijdlcqlimit}
\begin{align}
    \alpha' K_{\text{L}i} \cdot K_{\text{L}j} & = - K_{0,\,i} \, K_{0,\,j} + \lr \frac{\tilde{n}_i}{\tilde{R}} - \frac{\tilde{w}_i \tilde{R}}{\alpha'} \rr \lr \frac{\tilde{n}_j}{\tilde{R}} - \frac{\tilde{w}_j \tilde{R}}{\alpha'} \rr + k_i \cdot k_j \\[2pt]
    \alpha' K_{\text{R}i} \cdot K_{\text{R}j} & = - K_{0,\,i} \, K_{0,\,j} + \lr \frac{\tilde{n}_i}{\tilde{R}} + \frac{\tilde{w}_i \tilde{R}}{\alpha'} \rr \lr \frac{\tilde{n}_j}{\tilde{R}} + \frac{\tilde{w}_j \tilde{R}}{\alpha'} \rr + k_i \cdot k_j\,,
\end{align}
\end{subequations}
which, under the expansion of $K_0$ with respect to a large $\gamma$ in \eqref{eq:fes}, reduces to \eqref{eq:kllrr} after performing the double scaling limit and then replacing $\tilde{R} \rightarrow \alpha'_{\text{eff}} / R$\,, $\tilde{n}_i \rightarrow w_i$\,, $\tilde{w}_i \rightarrow n_i$ and $\alpha' \rightarrow \alpha'_\text{eff}$\,. This shows that the DLCQ amplitude is identical to the nonrelativistic closed string amplitude \eqref{eq:mcn}.

The same limit can also be applied to open string states. We consider open strings in a spacetime-filling D-brane background, with Neumann boundary conditions applied in all directions. Consider the component $A_1$ of a constant gauge potential that lies along $X^1$, associated with $U(m)$ Chan-Paton factors of the open strings. By a gauge transformation, we diagonalize $A_1$ such that it is in the Cartan subalgebra $U(1)^m$, with
\be
    A_1 = - \frac{1}{2\pi R_0} \, \text{diag} \left( \theta_1\,, \cdots\,, \theta_m \right).
\ee
The open string in the Chan-Paton state $|ij\rangle$ has the dispersion relation
\be
    K_0^2 - K_{A'} K_{A'} = \frac{\left( 2 \pi \, \tilde{n} + \theta_i - \theta_j \right)^2}{4 \pi^2 R_0^2} + \frac{N - 1}{\alpha'}\,.
\ee
States with a finite energy $E$ in the double-scaling limit are now associated with
\be \label{eq:ce}
    K_0 - \frac{2 \pi \, \tilde{n} + \theta_i - \theta_j}{2 \pi R_0} = \frac{\varepsilon}{2 \, \gamma} + O(\gamma^{-2})\,.
\ee
Therefore, in the double-scaling limit, the DLCQ dispersion relation is
\be \label{eq:dlcqdros}
    \varepsilon = \frac{\pi \, \tilde{R}}{2 \pi \, \tilde{n} + \theta_i - \theta_j} \left[ K_{A'} K_{A'} + \frac{1}{\alpha'} \left( N - 1 \right) \right].
\ee

To describe the exotic physics in the target-space with a lightlike compactification, it is useful to pass on to the T-dual frame \cite{Bergshoeff:2018yvt}. For this purpose, we rewrite the string action \eqref{eq:srel} in terms of the lightcone coordinates that satisfy the boundary conditions in \eqref{eq:yp}, such that
\be \label{eq:srellc}
    S_\text{rel.} = \frac{1}{4\pi\alpha'} \int d^2 \sigma \left( 2 \, \p_\alpha Y' \, \p^\alpha \overline{Y}{}' + \p_\alpha Y^{A'} \, \p^\alpha Y^{A'} \right). 
\ee
Recall that $Y'$ is compactified over a circle of radius $\tilde{R}$\,. We will focus on the closed string case, but generalizations to worldsheets with boundaries are straightforward and can be found in \cite{Gomis:2020fui}. The action \eqref{eq:srellc} is equivalent to the so-called parent action,
\be \label{eq:sp}
    S_\text{parent} = \frac{1}{4\pi\alpha'} \int_\Sigma d^2 \sigma \left( \p_\alpha Y^{A'} \, \p^\alpha Y^{A'} + 2 \, V_\alpha \, \p^\alpha \overline{Y}{}' + i \, \epsilon^{\alpha\beta} \, X^1 \, \p_\alpha V_\beta \right),
\ee
upon integrating out the auxiliary field $X^1$. Alternatively, we rewrite \eqref{eq:sp} as
\be
    S_\text{nonrel.} = \frac{1}{4\pi\alpha'} \int_\Sigma d^2 \sigma \left[ \p_\alpha Y^{A'} \, \p^\alpha Y^{A'} + \lambda \, \bar{\p} \lr \overline{Y}{}' + X^1 \rr + \bar{\lambda} \, \p \lr \overline{Y}{}' - X^1 \rr \right],
\ee
where we defined $\lambda = i V_\tau + V_\sigma$ and $\bar{\lambda} = - i V_\tau + V_\sigma$\,. Identifying $Y^{A'} = X^{A'}$ and $\overline{Y}{}' = X^0$\,, this becomes the string action for nonrelativistic string theory in \eqref{eq:nrsa}, where the T-dual circle is spacelike. In this sense, nonrelativistic string theory provides a first principle definition of the DLCQ of relativistic string theory via a T-duality along a spacelike circle. 

In the closed string section, this duality relation between nonrelativistic string theory and the DLCQ of relativistic string theory is also made manifest by T-dualizing the DLCQ dispersion relation \eqref{eq:dlcqdr}. This is done by using the maps $\tilde{R} = \alpha' / R$ and $\tilde{n} = w$\,, followed by replacing $\alpha'$ with $\alpha'_\text{eff}$\,. As a result, we recover the dispersion relation \eqref{eq:drcs} for nonrelativistic closed string theory on the T-dual side. The same T-dual relation also holds for the open string dispersion relation \eqref{eq:dlcqdros} in the DLCQ and the nonrelativistic open string dispersion relation \eqref{eq:nrosdr}. 

In the usual treatment of the DLCQ of string theory, no winding in the lightlike circle is considered. In general, however, a nonzero winding number along the lightlike circle can be introduced. In the T-dual frame, this is mapped to a KK number in $X^1$\,. Taking the T-dual of the nonrelativistic KLT relation \eqref{eq:nrklt}, we therefore obtain a general KLT factorization of amplitudes of the DLCQ of closed strings into the DLCQ of open strings on a stack of spacetime-filling D-brane. Such a T-duality transformation replaces the momentum number $n_i$ and the string winding number $w_i$ in the compactified $X^1$ direction with $\tilde{w}_i$ and $\tilde{n}_i$\,, respectively. Moreover, on the open string side, the coinciding D-branes carry the electric potentials in the same way as specified in \eqref{eq:vsfinal}.   

\section{Conclusions} \label{sec:concl}

In this paper, we studied various novel KLT relations in both relativistic and nonrelativistic string theories. Despite intense studies of KLT relations in string theory and QFTs, the introduction of winding states still brings new twists to the KLT relations, where the open strings are required to end on intriguing D-brane configurations that are determined by the closed string data. This suggests that a richer structure of winding string amplitudes still awaits to be discovered, which may eventually constitute essential ingredients for analyzing string compactification, a subject that is important for understanding the extended nature of strings. Moreover, the analogs of KLT relations in nonrelativistic string theory provide useful simplifications for further exploration of nonrelativistic closed string amplitudes. This is among the first steps towards generalizing the modern S-matrix program developed for relativistic string theory and QFTs to nonrelativistic physics. 

\acknowledgments
It is a pleasure to thank Jaume Gomis for many helpful discussions and collaborations. This research is supported in part by Perimeter Institute for Theoretical Physics. Research at Perimeter Institute is supported in part by the Government of Canada through the Department of Innovation, Science and Economic Development Canada and by the Province of Ontario through the Ministry of Colleges and Universities. Nordita is supported in part by NordForsk.

\newpage

\appendix

\section{One-Loop Amplitudes in Nonrelativistic Open String Theory} \label{app:ola}

Loop amplitudes in nonrelativistic bosonic closed string theory have been studied in \cite{Gomis:2000bd}, where it is shown that, as in relativistic string theory, there exists a standard perturbative expansion with respect to the genera of worldsheet Riemann surfaces. At one-loop order, integrating out the zero modes of the one-form field $\lambda$ in the path integral constrains $X$ to be a holomorphic map from the worldsheet to the longitudinal sector of the target space. For example, in the bosonic one-loop free energy at a finite temperature, \emph{i.e.}, the thermodynamic partition function of free closed strings, the path integral over the zero modes of $\lambda$ (and $\bar{\lambda}$) restricts the moduli space to a set of discrete points in the fundamental domain of $\text{SL}(2,\mathbb{C})$\,. Similar constraints are also generalized to higher-loops and general $\CN$-point nonrelativistic string amplitudes, restricting the moduli space to a lower dimensional submanifold \cite{Gomis:2000bd}. Nonrelativistic closed string amplitudes have been shown to match relativistic string amplitudes in the DLCQ \cite{Danielsson:2000gi, Bilal:1998vq}. The finiteness of the DLCQ string amplitudes now become manifestly true when posed in the T-dual language of nonrelativistic string theory. 

However, nonrelativistic open string amplitudes have not yet been studied in the literature.~\footnote{For amplitudes of NCOS, see \cite{Klebanov:2000pp} for example.} In \S\ref{sec:nos}, we gave the prescriptions for evaluating nonrelativistic open string amplitudes in general, and then in \S\ref{sec:kltfncsa} we considered tree-level nonrelativistic open string amplitudes in the context of KLT relations. In this section, we will employ our techniques developed through the evaluation of tree-level amplitudes in nonrelativistic string theory to one-loop bosonic open string amplitudes, 
and show that similar localization theorems in the moduli space arise. We will also compare these results derived in nonrelativistic string theory with relativistic string amplitudes in the DLCQ.
We start with the evaluation of the one-loop planar amplitude for the scattering of $\CN$ open string tachyons that end on D-branes transverse to the compactified $X^1$ circle. Then, we compute the one-loop free energy that represents the ensemble of free open string states at a finite temperature.

\subsection{\texorpdfstring{$\CN$}{N}-Point Planar Scattering for Tachyons} 

\begin{figure}[t!]
    \centering
    \begin{tikzpicture}
    \draw[decoration={markings, mark=at position .5 with
 {\arrow{>}}},postaction={decorate},line width = .3mm] (-1,-1)--(-1,1.5);
    \draw[line width = .3mm] (2,-1)--(2,2);
    \draw[line width = .3mm] (-2,-1)--(3,-1);
    \draw (2,2.2) node{$\sigma$};
    \draw (3.3,-1) node{$\tau$};
    \draw[line width = .3mm] (-1,1.5)--(2,1.5);
    \draw[decoration={markings, mark=at position .5 with
 {\arrow{<}}},postaction={decorate},line width = .3mm] (2,1.5)--(2,-1);
    \draw (1.7,-1) node {$\times$};
    \draw (.8,-1) node {$\times$};
    \draw (-1,-1) node {$\times$};
    \draw (2.25,1.5) node {$\pi$};
    \draw (-1.,-1.3) node {\scalebox{.85}{$\tau_\CN$}};
    \draw (-.1,-.8) node {\scalebox{2}{$\ldots$}};
    \draw (1.7,-1.3) node {\scalebox{.85}{$\tau_1$}};
    \draw (.8,-1.3) node {\scalebox{.85}{$\tau_2$}};
    \draw[decoration={markings, mark=at position .999 with
 {\arrow{>}}}, postaction={decorate},line width =.4 mm]
    (4,0)--(5,0);
    \draw (4.5,.3) node{\scalebox{.85}{$\rho^{}_i = e^{\tau^{}_i}$}};
    \draw[line width = .3mm] (5+2-1,-1)--(14-1,-1);
    \draw [decoration={markings, mark=at position .5 with
 {\arrow{<}}},postaction={decorate},line width= .4mm,domain=0:180] plot ({.7*cos(\x)+7.5+3-1}, {.7*sin(\x)-1});
    \draw [decoration={markings, mark=at position .5 with
 {\arrow{<}}},postaction={decorate},line width= .4mm,domain=0:180] plot ({3*cos(\x)+7.5+3-1}, {3*sin(\x)-1});
     \draw (12.5-1,-1) node {$\times$};
    \draw (13.2-1,-1) node {$\times$};
    \draw (10.2,-1) node {$\times$};
    \draw (13.6-1,-1.26) node{\scalebox{.8}{1}};
    \draw (11.85-1,-.8) node{\scalebox{2}{$\ldots$}};
    \draw (11.2-1,-1.3) node{\scalebox{.85}{$\rho^{}_\CN$}};
    \draw (13.2-1,-1.3) node {\scalebox{.85}{$\rho^{}_1$}};
    \draw (12.5-1,-1.3) node {\scalebox{.85}{$\rho^{}_2$}};
    \end{tikzpicture}
    \caption{The relation $\rho^{}_i = e^{\tau^{}_i}$\,, $i = 1, \cdots, \CN$ maps the strip it to the part of the annulus in the upper half plane.
    The vertex operators lying on the strip at $\sigma=0$ are mapped to the right side of the annulus. Here, $\tau_\CN = - 2 \pi \, t$ and $\rho^{}_\CN = e^{-2\pi t}$ are fixed. See \cite{Green:2012pqa}.}
    \label{fig:striptoannulus}
\end{figure}
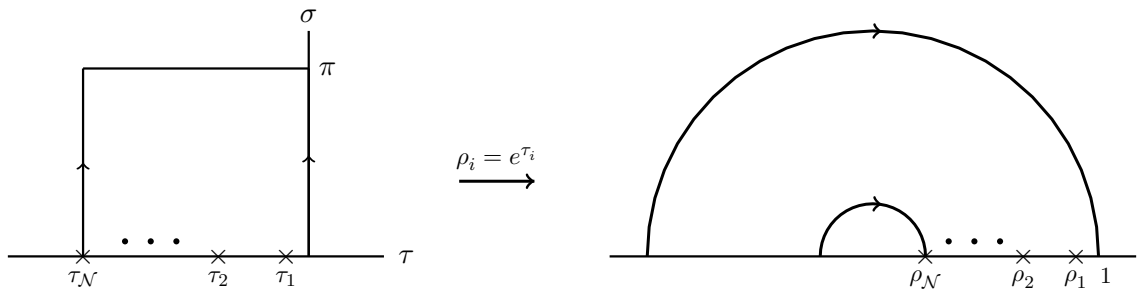

At one loop, the open string worldsheet is parametrized as a cylinder, given by region
\be\label{cylinderregion}
    0 \leq \sigma \leq \pi\,,
        \qquad
    \tau \sim \tau + 2 \pi \, t\,,
\ee
with $\tau$ the Euclidean time. The cylinder is formed by identifying the ends at $\sigma = 0\,, \pi$\,. Here, $t \in \mathbb{R}^+$ is the modulus. In terms of the complex coordinate $w = \sigma + i \, \tau$\,, the cylinder is a rectangle in the complex plane. We take this rectangular region to be $0 \leq \sigma \leq \pi$ and $- 2\pi \, t \leq \tau \leq 0$ in Fig.~\ref{fig:striptoannulus}, with the edges parallel to the $\sigma$-axis identified. Consider the radial coordinates with $z = e^{\tau + i \, \sigma}$\,, which maps the rectangular region to be the upper half of an annulus as in Fig.~\ref{fig:striptoannulus}.  

We now consider a planar diagram for an $\CN$-point open string scattering process, by inserting $\CN$ vertex operators on the same boundary of the cylinder at $\sigma = 0$ as in Fig.~\ref{fig:striptoannulus}. On the cylinder, there is a conformal Killing vector that is the translation parallel to the boundary. Using this symmetry we fix the $\CN$-th vertex operator to be at the bottom left corner of the rectangle, with $\sigma = 0$ and $\tau^{}_\CN = - 2\pi\,t$\,. Other vertex operators are located at $\tau_1^{}, \cdots, \tau_{\CN-1}^{}$ along the $\tau$-axis. The open strings are required to end on D-branes that are transverse to the $X^1$ circle. Therefore, boundaries of the worldsheet are joined to such D-brane configurations, satisfying Dirichlet boundary conditions.
Mapping to the half annulus, the vertex operators are located at 
$\rho^{}_i = e^{\tau^{}_i}$\,, with $i = 1, \cdots, \, \CN$ and $\rho^{}_\CN = e^{-2\pi t}$\,. 

\subsubsection*{Nonrelativistic open string amplitudes}

In nonrelativistic open string theory, since the open string OPEs in \eqref{eq:opeff} and the tachyonic vertex operators in \eqref{eq:vowsim} are in form the same as in relativistic string theory, the string amplitudes also take the same form as the ones for scattering between relativistic string tachyons,~\footnote{We follow \cite{Green:2012pqa} for one-loop planar amplitudes of $\CN$ relativistic open strings. If one replaces $\CI$ in \eqref{eq:defi} by 
\be
    \CI = \int d^{26} K \, \exp \! \ls - 2 \, \pi \, \alpha'_\text{eff} \, t \lr K + \frac{\sum_i K_i \, \ln \rho_i}{2 \, \pi \, t} \rr \rs = \lr \frac{1}{2 \, \alpha'_\text{eff} \, t} \rr^{\!13}\,,
\ee
then $\CA_\text{1-loop}^{(\CN)}$ becomes identical to Eq.~(8.1.63) in \cite{Green:2012pqa}. In Eq.~(8.1.63), $w = e^{-2\pi t}$ in our terminology, and the relation between $f(w)$ therein and the Jacobi theta function $\theta_1 (\nu|\tau)$ is given in Eq.~(8.A.26) in \cite{Green:2012pqa}.}
\begin{align}\label{eq:applanar1loop}
    \CA^{(\CN)}_\text{1-loop} = g_\text{o}^{(\CN)} \int_0^\infty \! dt \, \bigl[ \eta(it) \bigr]^{-24} \int_0^1 \, \prod_{i=1}^{\CN-1} \frac{d\rho_i}{\rho_i} \, \Theta(\rho_i - \rho_{i+1}) \, \prod_{j<k} \psi_{jk}^{2 \, \alpha'_\text{eff} K_j \cdot K_k} \, \CI\,,
\end{align}
where
\begin{subequations}
\begin{align}
    \psi_{ij} & = -2 \pi \, i \, \exp \bigl( \pi \, t^{-1} \, \nu_{ij}^2 \bigr) \frac{\theta_1 \! \lr \nu_{ij} \, | \, i t \rr}{\theta'_1 \! \lr 0 \, | \, i t \rr}\,, \\[2pt]
    \CI & = \sum_w \int d\varepsilon \, d^{24} k^{A'} \exp \! \ls - 2 \, \pi \, \alpha'_\text{eff} \, t \lr K + \frac{\sum_{i=1}^\CN K_i \ln \rho_i}{2\pi \, t} \rr^{\!\!2\,} \rs. \label{eq:defi}
\end{align}
\end{subequations}
Here, $\nu_{ij} = \ln \lr \rho_i / \rho_j \rr / (2\pi i)$\,; $K$ and $K_i$ are given in \eqref{eq:klrW}, with
\be \label{eq:kki}
    K = \lr \frac{\varepsilon}{2} + \frac{W R}{\alpha'_\text{eff}}\,,\, \frac{\varepsilon}{2} - \frac{W R}{\alpha'_\text{eff}}\,, k_{A'} \rr,
        \qquad%
    K_i = \lr \frac{\varepsilon_i}{2} + \frac{W_i R}{\alpha'_\text{eff}}\,,\, \frac{\varepsilon_i}{2} - \frac{W_i R}{\alpha'_\text{eff}}\,, k^{A'}_i \rr,
\ee
and $W$ and $W_i$ the total winding numbers for open strings ending on D-branes that are transverse to the compactified $X^1$ circle. Note that only the integer winding $w$ is summed over.
The notation $\Theta (\rho_i - \rho_{i+1})$ is the Heaviside function that specifies the cyclic ordering of the open string vertex operators on the boundary of the cylinder at $\sigma = 0$\,. Moreover, $\theta_1 (\nu|\tau)$ is a Jacobi theta function and $\theta_1' (\nu|\tau) = \p_\nu \theta_1 (\nu|\tau)$\,.
Using \eqref{eq:kki}, we find
\be \label{eq:kij}
    \alpha'_\text{eff} \, K_i \cdot K_j = - (\varepsilon_i \, W_j + \varepsilon_j \, W_i) \, R + \alpha'_\text{eff} \, k_i \cdot k_j\,,
\ee
and 
\be
    \CI = \frac{1}{(2 \, \alpha'_\text{eff} \, t)^{12}} \sum_{w} \int d\varepsilon \, \exp \! \ls 4 \pi R \lr \varepsilon + \frac{\sum_i \varepsilon_i \ln \rho_i}{2\pi t} \rr \lr W \, t + \frac{\sum_j W_j \ln \rho_j}{2 \pi} \rr \rs.
\ee
In order for the integral over $\varepsilon$ to be well defined, we promote this integral to be over the complex plane, followed by performing a Wick rotation as in \cite{Bilal:1998vq}. Integrating over $\varepsilon$\,, we find
\be \label{eq:inrst}
    \CI \rightarrow \frac{\pi}{(2 \, \alpha'_\text{eff} \, t)^{12}} \sum_w \frac{1}{R} \, \delta \Bigl( 2 \pi t \, W + \sum_i W_i \, \ln \rho_i \Bigr). 
\ee
Therefore, 
\begin{align} \label{eq:anosol}
\begin{split}
    \CA^{(\CN)}_\text{1-loop} & = \frac{\pi \, g_\text{o}^{(\CN)}}{R}
    \int_0^\infty \frac{dt}{(2 \, \alpha'_\text{eff} \, t)^{12}} \, \bigl[ \eta(it) \bigr]^{-24} \int_0^1 \prod_{i=1}^{\CN-1} \frac{d\rho_i}{\rho_i} \, \Theta(\rho_i - \rho_{i+1}) \\[2pt] 
    & \hspace{.9cm} \times \prod_{j<k} \psi_{jk}^{2 \, \alpha'_\text{eff} \,  k_j \cdot k_k - 2 \, (\varepsilon_j \, W_k + \varepsilon_k \, W_j) R} \, \sum_w \delta \Bigl( 2\pi t \, W + \sum_\ell W_\ell \, \ln \rho_\ell \Bigr).
\end{split}
\end{align}
The Dirac delta function constrains the moduli space to be a submanifold that satisfies the condition
\be
    2 \pi t \, W + \sum_{i=1}^{\CN} W_i \, \ln \rho_i = 0\,. 
\ee

\subsubsection*{Zero $\alpha'$ limit of relativistic string amplitudes}

The same result in \eqref{eq:anosol} can be reproduced by taking a zero $\alpha'$ limit of winding string amplitudes in relativistic open string theory, in the presence of an electric $B$-field. This zero $\alpha'$ limit has been reviewed in \S\ref{sec:zalwkltr}. Consider string amplitudes for scatterings between open strings in relativistic string theory with the background field configuration specified in \eqref{eq:hgb2}. The open strings are required to end on D-branes that are localized in the compactified $X^1$ direction. These D-branes extend in all the other directions. The vertex operator for a tachyonic open string is
\be
    \CV_\text{open} = g_\text{o}\, :\! \exp \bigl( i K_\text{L} \cdot X_\text{L} + i K_\text{R} \cdot X_\text{R} \bigr) \!:
\ee
with $K_\text{L,\,R}$ defined similarly as in \eqref{eq:krosb} but now in the critical background field configuration specified in \eqref{eq:hgb2},~\footnote{Only $K_\text{R}$ is given in \eqref{eq:krosb}, from which $K_\text{L}$ can be obtained by flipping the signs in front of the total winding number in the second entry.} with
\begin{subequations} \label{eq:klkr22}
\begin{align} 
    K_{\text{L}\mu} & = \lr \varepsilon - \frac{W R}{\alpha'} \, B\,,\, - \frac{W R}{\alpha'}\,,\, \sqrt{\frac{\alpha'_\text{eff}}{\alpha'}} \, k_{A'} \rr, \\[2pt]
    K_{\text{R}\mu} & = \lr \varepsilon - \frac{W R}{\alpha'} \, B\,,\, \frac{WR}{\alpha'}\,,\, \sqrt{\frac{\alpha'_\text{eff}}{\alpha'}} \, k_{A'} \rr.
\end{align}
\end{subequations}
Here, $\alpha'$ is the Regge slope in relativistic string theory, and $W$ refers to the total open string winding number. 
Moreover, 
$X^\mu_{\text{L},\,\text{R}}$ are (anti-)holomorphic coordinates. We impose Neumann boundary conditions in $X^{0,A'}$ and Dirichlet boundary condition in $X^1$\,. In the critical field configuration, we set $B = -1$ as in \S\ref{sec:zalwkltr}. The associated open string amplitude is
\begin{align} \label{eq:AN1looprel}
    \tilde{\CA}^{(\CN)}_\text{1-loop} = \tilde{g}_\text{o}^{(\CN)} \int_0^\infty \! dt \, \bigl[ \eta(it) \bigr]^{-24} \int_0^1 \, \prod_{i=1}^{\CN-1} \frac{d\rho_i}{\rho_i} \, \theta(\rho_i - \rho_{i+1}) \prod_{j<k} \psi_{jk}^{\alpha' \bigl( K_{\text{L}j} \cdot K_{\text{L}k} + K_{\text{R}j} \cdot K_{\text{R}k} \bigr)} \, \tilde{\CI}\,,
\end{align}
where
\begin{align} \label{eq:ip}
\begin{split}
    \tilde{\CI} = \sum_w \int d\varepsilon \, d^{24} k^{A'} \, & \exp \ls - \alpha' \pi \, t \lr K_\text{L} + \frac{\sum_i K_{\text{L}i} \, \ln \rho_i}{2\pi t} \rr^{\!2} \, \rs \\[2pt]
    \times & \exp \ls - \alpha' \pi \, t \lr K_\text{R} + \frac{\sum_j K_{\text{R}j} \, \ln \rho_j}{2\pi t} \rr^{\!2} \, \rs.
\end{split}
\end{align}
Performing a Wick rotation for the energy $\varepsilon$\,, we are able evaluate the integrals,
\be
    \tilde{\CI} \rightarrow \frac{1}{(2\,\alpha'_\text{eff}\,t)^{12}} \sum_w \frac{1}{\sqrt{2 \, \alpha' \, t}} \, \exp \! \ls - \frac{R^2}{{2\pi \alpha'} \, t} \Bigl( 2\pi t \, W + \sum_i W_i \, \ln \rho_i \Bigr)^{2} \rs
\ee
Applying the limit $\alpha' \rightarrow 0$\,, and noting that
\be \label{eq:sigmazero}
    \lim_{\sigma \rightarrow 0} \frac{1}{\sqrt{2\pi} \sigma} \, \exp \lr - \frac{\chi^2}{2 \, \sigma^2} \rr = \delta ( \chi )\,,
\ee
and
\be
    \alpha' \bigl( K_{\text{L}i} \cdot K_{\text{L}j} + K_{\text{R}i} \cdot K_{\text{R}j} \bigr) \rightarrow - 2 \, \bigl( \varepsilon_i \, W_j + \varepsilon_j \, W_i\bigr) R + \alpha'_\text{eff} \, k_{i} \cdot k_{j}
\ee
we find that $\tilde{\CI}$ becomes $\CI$ in \eqref{eq:inrst} and $\tilde{\CA}^{\CN}_\text{1-loop}$ becomes $\CA_\text{1-loop}^{(\CN)}$ in \eqref{eq:anosol}.
Therefore, the nonrelativistic open string amplitude \eqref{eq:anosol} is reproduced in the $\hat{\alpha}' \rightarrow 0$ limit.  

\subsubsection*{DLCQ Amplitudes from an infinite boost limit}

In the T-dual picture, the theory is described by the DLCQ of relativistic string theory on a compactification over a lightlike circle of radius $\tilde{R}$\,. As discussed in \S\ref{sec:stdlcq}, this is usually defined by starting with relativistic string theory on a spacelike circle followed by performing a subtle infinite boost limit \cite{Seiberg:1997ad}. We therefore first consider open string amplitudes on spacetime-filling D-branes with the spatial direction $Y^1$ compactified over a circle of radius $R_0$\,, with $Y^1$ satisfying the periodic boundary condition in \eqref{eq:y1pbc}. The double scaling limit in \S\ref{sec:stdlcq} requires $\tilde{R} = 2 \gamma R_0$\,, where we hold $\tilde{R}$ fixed while taking the limits $\gamma \rightarrow \infty$ and $R_0 \rightarrow 0$\,.  
Consider open string tachyon insertions all on one boundary of the worldsheet cylinder. These open string tachyon states carry quantized momenta $p^{}_1 = \tilde{n} / R_0$\,, $\tilde{n} \in \mathbb{Z}$ along the $Y^1$ direction. The resulting $\CN$-point amplitude takes the same form of \eqref{eq:applanar1loop}, but the kinematic data now becomes
\be
    K = \left( K_0\,, \, \frac{\tilde{n}}{R_0}\,, \, k_{A'} \right)\,,
        \qquad%
    K_i = \left( K_{0,\,i}\,, \, \frac{\tilde{n}_i}{R_0}\,, \, k_{A'} \right)\,.
\ee
The effective energies $\varepsilon$ and $\varepsilon_i$  are defined in \eqref{eq:fes}, with
\be
    K_0 - \frac{\tilde{n}}{R_0} = \frac{\varepsilon}{2 \gamma} + O(\gamma^{-2})\,,
        \qquad%
    K_{0,\,i} - \frac{\tilde{n}_i}{R_0} = \frac{\varepsilon_i}{2 \gamma} + O(\gamma^{-2})\,.
\ee
We also have to replace the factor in \eqref{eq:applanar1loop} with
\be
    \CI' = \sum_{\tilde{n}} \int d\varepsilon \, d^{24} k^{A'} \, \exp \! \ls - 2 \, \pi \, \alpha' \lr K + \frac{\sum_{i=1}^\CN K_i \, \ln \rho_i}{2 \pi \, t} \rr^{\!2} \rs
\ee
Consequently, performing a Wick rotation for the intermediate energy $\varepsilon$\,, we find 
\be
    \CI' \rightarrow \frac{\tilde{R}}{(2\alpha' t)^{25/2}} \sum_{\tilde{n}} \frac{1}{R_0} \exp \! \ls - \frac{\alpha'}{2 \pi R_0^2 \, t} \Bigl( 2 \pi t \, \tilde{n} + \sum_i \tilde{n}_i \, \ln \rho_i \Bigr)^2 \rs\,. 
\ee
Using \eqref{eq:sigmazero}, we perform the double scaling limit by sending $R_0 \rightarrow 0$ while holding $\tilde{R}$ fixed,
\be \label{eq:idlcq}
    \CI' \rightarrow \frac{\pi}{(2\alpha' \, t)^{12}} \sum_{\tilde{n}} \frac{\tilde{R}}{\alpha'} \, \delta \Bigl( 2\pi t \, \tilde{n} + \sum_{i} \tilde{n}_i \, \ln \rho_i \Bigr)\,.
\ee
We already learned in \S\ref{sec:stdlcq} that the DLCQ of relativistic string theory related to nonrelativistic string theory via a T-duality transformation.
Identifying $\alpha'$ with $\alpha'_\text{eff}$\,, and plugging in the T-dual relations $\tilde{R} = \alpha' / R$ and $\tilde{n} = w$ into \eqref{eq:idlcq}, we find that $\CI'$ becomes $\CI$ in \eqref{eq:inrst} with $W_i = w_i$\,. The fractional part of open string windings in \eqref{eq:inrst} can also be recovered by including associated Wilson lines on the DLCQ side. Moreover, similar to the discussion around \eqref{eq:kijdlcqlimit} for closed strings, the kinematic variable $K_i \cdot K_j$ reduces to the desired one in \eqref{eq:kij} after performing the double scaling limit followed by plugging in the T-dual relations. We therefore obtain the same one-loop string amplitude \eqref{eq:anosol} for nonrelativistic open strings from the DLCQ of relativistic string theory.

\subsection{Free Energy} 

In order to compute the one-loop free energy at the inverse temperature $\beta = 1 / T$ in nonrelativistic open string theory, we first perform a Wick rotation of the target-space time direction from $X^0$ to $X^0_\text{E} = i \, X^0$\,, and then compactify $X^0_\text{E}$ over an imaginary timelike circle of period $\beta$, with $X^0_\text{E} \sim X^0_\text{E} + \beta$\,. In addition, as in \S\ref{sec:kltfnsa}, we also compactify the longitudinal spatial direction $X^1$ over a circle of radius $R$\,. These lead to the periodic boundary conditions,
\begin{subequations}
\begin{align}
    X^0_\text{E} (\sigma\,, \tau+2\pi t) &= X^0 + m \beta\,,
        &%
    X^1(\sigma\,, \tau+2\pi t) &= X^1\,, \\[2pt]
    X^0_\text{E} (\sigma + \pi\,, \tau) &= X^0\,,
        &%
    X^1(\sigma + \pi\,, \tau) &= X^1 - 2 \pi R \, W\,.
\end{align}
\end{subequations}
Here, $m \in \mathbb{Z}$ and $W$ contains an integer part $w \in \mathbb{Z}$\,. The boundaries of the worldsheet cylinder is attached to D-branes that are transverse to $X^1$\,, while the worldsheet imaginary time direction is wrapped around the Euclidean time direction $X^0_\text{E}$ in the target space. Together with the Dirichlet boundary condition $\p_\tau X^1 = 0$ and the Neumann boundary conditions $\p_\sigma X^0 = 0$ at $\sigma = 0, \pi$\,, we find the following zero modes for the worldsheet fields $X^A$:
\be
    X^0_\text{E} = x^0_\text{E} + \frac{m \beta}{4\pi t} \log (z \bar{z}) + \text{oscillations}\,,
        \qquad%
    X^1 = x^1 + i\,W R \, \log\lr\frac{z}{\bar{z}}\rr + \text{oscillations}\,.
\ee
Using \eqref{eq:qqb}, we find the eigenvalues
\be
    q = - \bar{q} = \frac{W R}{\alpha'_\text{eff}} - \frac{m \beta}{4\pi \alpha'_\text{eff} \, t}\,.
\ee
Moreover, since the momentum in $X^1$ is zero, we have $p = \bar{p} = i \, \varepsilon/2$\,, where $\varepsilon$ is the ``momentum" conjugate to $X^0_\text{E}$ and is quantized as $\varepsilon = 2 \pi s / \beta$\,.
The one-loop free energy for nonrelativistic string theory can be computed by borrowing the analogous expression for relativistic string theory and then plugging in the kinematic data defined in \eqref{eq:qn}, with
\be \label{eq:kdfe}
    K_\mu = K_{\text{L}\mu} = K_{\text{R}\mu} = \lr \frac{i \, \pi s}{2 \beta} + \frac{WR}{\alpha'_\text{eff}} - \frac{m\beta}{4\pi\alpha'_\text{eff} \, t}\,,\, \frac{i \, \pi s}{2 \beta} - \frac{WR}{\alpha'_\text{eff}} + \frac{m\beta}{4\pi\alpha'_\text{eff} \, t}\,,\, k_{A'} \rr.
\ee

To proceed, we first recall the one-loop vacuum amplitude for relativistic open string theory, where a general physical state takes the form \cite{Polchinski:1998rq}
\be
    | N\,; K_\mu \rangle = \ls \prod_{A'=2}^{25} \prod_{n=1}^\infty c^{-1}_{A'\!,\,n} \, \bigl( \alpha^{A'}_{-n} \bigr)^{N_{A'\!, \, n}} \rs |0; K_\mu \rangle\,,  
        \qquad%
    c_{A'\!, \,n} = \sqrt{n^{N_{A', n}} \, N_{A'\!, n}!}\,.
\ee
Here, $N_{A'\!,\, n}$ denotes the occupation number for the mode $(A'\!, \, n)$ in the lightcone quantization, and $N$ is the level of the state, with
\be 
    N = \sum_{A'=2}^{25} \sum_{n=1}^\infty n \, N_{A'\!, \, n}\,.
\ee
The state $|0; K_\mu\rangle$ is annihilated by any lowering operator $\alpha^{A'}_{n}$\,. Then, the vacuum planar amplitude from the worldsheet cylinder is
\begin{align} \label{eq:pffe}
    \mathcal{Z} = \int_0^\infty \frac{dt}{2t} \, \text{Tr}' \left( e^{-2\pi t \, H} \right) 
    = \int_0^\infty \frac{dt}{2t} \lr  \prod_{A'\!,\,n} \sum_{N_{A'\!,\,n}=0}^\infty e^{-2\pi t \, (n \, N_{A'\!,\,n}-1)} \rr \CI\,, 
\end{align}
where we denoted
the zero-mode contribution by
\begin{align} \label{eq:ci}
    \CI & = \sum_{m, w, s} \int d^{24}k_{A'} \, e^{-2 \pi \, \alpha'_\text{eff} \, t \, K^2}.
\end{align}
Note that we have introduced the open string Hamiltonian $H$, and the prime in the trace implies that we already omitted the unphysical states that are canceled by the ghosts. Moreover, since we do not have any vertex operator insertions, the conformal Killing group is no longer used to fix any position of the vertex operators. Instead, a measure $(2\,t)^{-1}$ has been introduced in \eqref{eq:pffe} so that the volume of the conformal Killing group is divided out. Performing the sum over $n$ and $N_{A'\!,\,n}$ in \eqref{eq:pffe} brings the vacuum amplitude to the same form as \eqref{eq:AN1looprel}, by rewriting the sum over all oscillating modes in terms of the Dedekind eta functions as
\be \label{eq:iddeta}
    \prod_{A'\!,\,n} \sum_{N_{A'\!,\,n}=0}^\infty e^{-2\pi t \, (n \, N_{A'\!,\,n}-1)} = \biggl[ e^{\pi t/12} \prod_{n=1}^\infty \lr 1- e^{-2\pi t \, n} \rr^{-1} \bigg]^{24} = \bigl[ \eta (it) \bigr]^{-24}\,.
\ee

To produce the desired vacuum amplitude in nonrelativistic string theory, we plug \eqref{eq:kdfe} into \eqref{eq:ci}. Then, the factor $\CI$ is evaluated to be
\be
    \CI = \sum_{m, w, s} \int d^{24} k_{A'} \, \exp \! \ls \pi i \lr \frac{4\pi R W}{\beta} \, t - m \rr s - 2 \pi \, \alpha'_\text{eff} \, t \, k^2 \rs.
\ee
By using a representation of the Dirac delta in terms of the Fourier series,
\be
    2\pi \delta \lr x \rr = \sum_s e^{i s x}\,,
\ee
we obtain
\be
    \CI = \sum_{m,w} \frac{\beta}{2\pi R W} \, \delta \! \lr t - \frac{\beta m}{4\pi R W} \rr \int {d^{24} k_{A'}} \, \exp \! \lr - \frac{\alpha'_\text{eff} \, \beta m \, k^2}{2RW} \rr.
\ee
Integrating over the modulus $t$ in \eqref{eq:pffe} yields
\be \label{eq:czaf}
    \mathcal{Z} = \int d^{24} k_{A'} \prod_{A'=2}^{25} \prod_{n=1}^\infty \sum_{w, m =1}^\infty \sum_{N_{A'\!,\,n} = 0}^\infty \frac{1}{m} \, \exp \lr - \beta \, m \frac{\alpha'_\text{eff} \, k^2 + n \, N_{A'\!,\,n} - 1}{2RW} \rr. 
\ee
Using the identity in \eqref{eq:iddeta}, we find that the vacuum amplitude takes the following form:
\begin{align}\label{eq:torusfreeenergy}
    \mathcal{Z} 
    & = \sum_{m, w} \frac{1}{m} \lr \frac{2\pi RW}{\alpha'_\text{eff} \, \beta m} \rr^{\!12} \ls \eta \! \lr \frac{i \beta m}{4\pi R W} \rr \rs^{-24}\!.
\end{align}
Alternatively, 
we perform the sum over $m$ in \eqref{eq:czaf} and compute the Helmholtz free energy as follows:
\be \label{eq:feff}
    \mathscr{F} = - T \mathcal{Z} = T \sum_{\varepsilon} \, D(\varepsilon) \, \ln \lr 1 - e^{-\beta \, \varepsilon} \rr,
        \qquad%
    \varepsilon = \frac{\alpha'_\text{eff}}{2 w R} \ls k_{A'} k_{A'} + \frac{1}{\alpha'_\text{eff}} \lr N - 1 \rr \rs.
\ee
Here, $\varepsilon$ takes the same form as in \eqref{eq:nrosdr}, which is the energy of a free open string.
We also used $D(\varepsilon)$ to denote the density of states associated with the energy $\varepsilon$\,. Manifestly, $\mathscr{F}$ in \eqref{eq:feff} is the free energy of free bosonic nonrelativistic open strings. This is the nonrelativistic open string analog of the result in \cite{Gomis:2000bd}, where the Hagedorn temperature that signals a phase transition at high energies is also given.

\newpage

\bibliographystyle{JHEP}
\bibliography{draft}
\end{document}